\UseRawInputEncoding
\documentclass[10pt,final,twocolumn]{IEEEtran}
\long\def\comment#1{}

\usepackage{textcomp}
\usepackage{cite}
\usepackage{graphicx}
\usepackage{subfigure}
\usepackage{psfrag}
\usepackage{url}
\usepackage{stfloats}
\usepackage{amsmath}
\usepackage{amssymb,amsfonts}
\usepackage{amsthm}
\usepackage{float}
\usepackage[usenames]{color}
\usepackage{algorithm,algorithmic}
\usepackage{multirow}
\usepackage{makecell}
\usepackage{threeparttable}

\newtheorem{definition}{Definition}

\newtheorem{example}{Example}

\begin{document}

\setlength{\arraycolsep}{0.3em}

\title{A Survey of Decision Making in Adversarial Games
\thanks{}}

\author{Xiuxian Li, Min Meng, Yiguang Hong, and Jie Chen
\thanks{This work was supported by National Natural Science Foundation of China under Grant 62003243, Grant 62103305 and Grant 62088101, Shanghai Municipal Commission of Science and Technology No. 19511132100, 19511132101, and Shanghai Municipal Science and Technology Major Project, No. 2021SHZDZX0100.}
\thanks{The authors are with the Department of Control Science and Engineering, College of Electronics and Information Engineering, and the Shanghai Research Institute for Intelligent Autonomous Systems, Tongji University, Shanghai, China (Email: xxli@ieee.org, mengmin@tongji.edu.cn, yghong@tongji.edu.cn, chenjie206@tognji.edu.cn).}
\thanks{X. Li and M. Meng are also with the Institute for Advanced Study, Tongji University, Shanghai, China.}
}

\maketitle

\setcounter{equation}{0}
\setcounter{figure}{0}
\setcounter{table}{0}

\begin{abstract}
Game theory has by now found numerous applications in various fields, including economics, industry, jurisprudence, and artificial intelligence, where each player only cares about its own interest in a noncooperative or cooperative manner, but without obvious malice to other players. However, in many practical applications, such as poker, chess, evader pursuing, drug interdiction, coast guard, cyber-security, and national defense, players often have apparently adversarial stances, that is, selfish actions of each player inevitably or intentionally inflict loss or wreak havoc on other players. Along this line, this paper provides a systematic survey on three main game models widely employed in adversarial games, i.e., zero-sum normal-form and extensive-form games, Stackelberg (security) games, zero-sum differential games, from an array of perspectives, including basic knowledge of game models, (approximate) equilibrium concepts, problem classifications, research frontiers, (approximate) optimal strategy seeking techniques, prevailing algorithms, and practical applications. Finally, promising future research directions are also discussed for relevant adversarial games.
\end{abstract}


\section{Introduction}\label{sec1}

Game theory has long been a powerful and conventional paradigm for modeling complex and intelligent interactions among a group of players and improving decision making for selfish players, since the seminal work \cite{von1947theory,nash1950equilibrium,nash1951non} by John von Neumann, John Nash, and others. Hitherto, it has found a vast range of real-world applications in a variety of domains, including economics, biology, finance, computer science, politics, and so forth, where each individual player is only concerned with its own interest \cite{fudenberg1991game,osborne1994course,bacsar2018handbook}. It played an extremely important role even during the Cold War in the 60s, and has been employed by many national institutions in defense, such as United States Agencies for security control \cite{aumann1995repeated}.

\begin{figure}[h]
\centering
\includegraphics[width=2.0in]{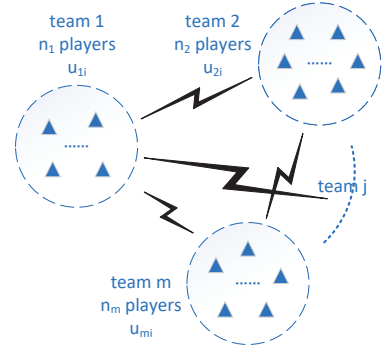}
\caption{A general framework of adversarial games with simultaneous or sequential moves, perfect or imperfect information, and symmetric or asymmetric information, where triangles denote players and there exist $m$ teams, within which team members play in a cooperative manner, while the play among teams is adversarial and usually zero-sum, i.e., $\sum_{i=1}^m\sum_{j=1}^{n_i}u_{ij}(x_{ij},x_{-ij})=0$ for all strategies with the subscript $ij$ representing the $j$-th player in $i$-th team whose strategy and utility function are denoted as $x_{ij}$ and $u_{ij}$, respectively. And $x_{-ij}$ is the strategy profile of all players except the $j$-th player in team $i$.}
\label{g_ag}
\end{figure}

Adversarial games are a class of particularly important game models, where players deliberately compete with each other while simultaneously achieving their own utility maximization. To date, adversarial games have been an orthodox framework for shaping high-efficient decision making in numerous realistic applications, such as poker, chess, evader pursuing, drug interdiction, coast guard, cyber-security, and national defense, etc. For example, in Texas Hold'em poker, which has been one of primary competitions as a benchmark for testing researchers' proposed algorithms in game theory and artificial intelligence (AI) held by international well-known conferences such as AAAI, multiple players compete against each other to win the game by seeking sophisticated strategy techniques \cite{bard2013annual}. Generally speaking, adversarial games enjoy several main features as follows: 1) hardness of efficient and fast algorithms design with limited computing resources and/or samples; 2) imperfect information for many practical problems, that is, some information is private to one or more players which, however, is hidden from other players, such as the card game of poker; 3) large models, including large action spaces and information sets, for example, the adversary space in the road networks security problem is of the order $10^{18}$ \cite{nguyen2016towards}; 4) incomplete information for a multitude of real-life applications, that is, one or more agents do not know what game is being played (e.g., the number of players, the strategies available to each player, and the payoff for each strategy), and in this case, the very game being played is generally represented with players' uncertainties, like uncertain payoff functions with uncertain parameters; and 5) possible dynamic trait, i.e., the played game is sometimes time-varying, instead of static, for example, a poacher may have different poaching strategies in a wildlife park as the environment varies with seasons. It is worth pointing out that incomplete information is understood distinctly from imperfect information here, as distinguished by some researchers, although they are interchangeably used in some literature. In addition, other possible characteristics include bounded rationality, where players may be not fully rational, such as arbitrarily random lone wolf attacks by terrorists. However, it is noteworthy that not all adversarial games are with imperfect and/or incomplete information, for example, the game of Go has both perfect and complete information, since it has explicit game rules and all chess pieces' positions are visible to both players at all times as well as the actions of the opponent, which has been well solved by well-known AI agents, such as AlphaGo and AlphaZero \cite{silver2016mastering,silver2017mastering,silver2018general}.

As the competitive feature is ubiquitous in a large number of real-world applications, adversarial games have been extensively investigated until now \cite{sinha2018stackelberg,li2021survey,sohrabi2020survey,wang2022cooperative,li2022survey,ho2022game}. For example, the authors in \cite{sinha2018stackelberg} provided a broad survey of technical advances in Stackelberg security games (SSG) in 2018, the authors in \cite{li2021survey} reviewed some main Nash equilibrium (NE) computing algorithms for extensive-form games with imperfect information based on counterfactual regret minimization (CFR) methods, the authors in \cite{sohrabi2020survey} reviewed a combined use of game theory and optimization algorithms along with a new categorization for researches conducted in this area, the authors in \cite{wang2022cooperative} reviewed distributed online optimization, federated optimization from the perspective of privacy-preserving mechanisms, and cooperative/non-cooperative games from two facets, i.e., minimizing global costs and minimizing individual costs, and the authors in \cite{li2022survey} surveyed recent advances of decentralized online learning, including decentralized online optimization and online game, from the perspectives of problem classifications, performance metrics, state-of-the-art performance results, and potential research directions in future. Additionally, in consideration of the importance of game theory in national defense, some reviews of game theory in defense applications were succinctly provided in \cite{ho2022game,shishika2020review}, and a survey of defensive deception based on game theory and machine learning (ML) approaches was presented in \cite{zhu2021survey}. Nonetheless, a thorough overview for adversarial games from the perspectives of the basic models' knowledge, equilibrium concepts, optimal strategy seeking techniques, research frontiers, and prevailing algorithms is still lacking.

Motivated by the above facts, this survey aims to provide a systematic review on adversarial games from several dimensions, including the models of three main models frequently employed in adversarial games (i.e., zero-sum normal-form and extensive-form games, Stackelberg (security) games, and zero-sum differential games), (approximate) optimal strategy concepts (i.e., NE, correlated equilibrium, coarse-correlated equilibrium, strong Stackelberg equilibrium, team-maxmin equilibrium, and corresponding approximate ones), (approximate) optimal strategy computing techniques (e.g., CFR methods, AI methods), state-of-the-art results, prevailing algorithms, potential applications, and promising future research directions. To the best of our knowledge, this survey is the first systematic overview on adversarial games, generally providing an orthogonal and complementary component for the aforementioned survey papers, which may aid researchers and practitioners in relevant domains. Please note that the three game models are not mutually exclusive, but may overlap for the same game from different viewpoints. For example, Stackelberg games and differential games can also be zero-sum games, etc. In addition, there actually exist other models leveraged for adversarial games, such as Bayesian games, Markov games (or stochastic games), signaling games, behavioral game theory and evolutionary game theory. However, we are not ambitious to review all of them in this survey, since each of them is of independent interest and pretty abundant in existing diverse materials.


The structure of this survey is organized as follows. The detailed game models and solution concepts are introduced in Section \ref{sec2}, the existing main literature is reviewed along with state-of-the-art results in Section \ref{sec3}, some prevailing algorithms are expounded in Section \ref{sec4}, an array of applications are presented in Section \ref{sec5}, promising future research directions are discussed in Section \ref{sec6}, and finally the conclusion is drawn in Section \ref{sec7}.

{\em Notations:} Define $[n]:=\{1,2,\ldots,n\}$ be the set of $n$ positive numbers for an integer $n$. Denote by $\mathbb{R}$, $\mathbb{R}^d$, and $\mathbb{R}_+^d$ the sets of real numbers, $d$-dimensional real vectors, and nonnegative $d$-dimensional real vectors, respectively. For a finite set $S$ with $m$ elements, define $\Delta(S)=\Delta^m:=\{x\in\mathbb{R}_+^m:\sum_{i=1}^m x_i=1\}$ (i.e., the simplex of dimension $m-1$), and $|S|$ be the cardinality of $S$. Let $\mathbb{P}$ and $\mathbb{E}$ denote the mathematical probability and expectation, respectively. Let $x^\top$ denote the transpose of $x$, and $\langle\cdot,\cdot\rangle$ be the inner product. ${\bf 0}$ and ${\bf 1}$ denote vectors or matrices of all entries $0$ and $1$ with compatible dimension in the context, respectively, sometimes with explicit subscript being the dimension.

\section{Models of Adversarial Games}\label{sec2}

This section provides three main models for adversarial games, i.e., zero-sum normal-form and extensive-form games, Stackelberg (security) games and differential games, along with solution concepts in these game models, and a general framework of adversarial games is illustrated in Fig. \ref{g_ag}.

\subsection{Zero-Sum Normal-Form and Extensive-Form Games}\label{subs2.1}

Normal-form and extensive-form games are two widely employed game models, accounting for simultaneous or sequential actions committed by the players in a game.

\begin{figure}[h]
\centering
\includegraphics[width=1.6in]{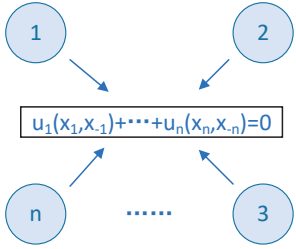}
\caption{A schematic illustration of zero-sum games with $n$ players.}
\label{ZSG}
\end{figure}

{\bf Normal-Form Games (NFGs).} A {\em normal-form} (or {\em strategic-form}) game is denoted by a tuple $(N,A,u)$ \cite{fudenberg1991game}, where $N:=[n]$ is a finite set of players. In the meantime, $A:=A_1\times\cdots\times A_n$ is the action profile set for all players, where $A_i$ is the set of pure actions or strategies available to player $i\in[n]$, and $a=(a_1,\ldots,a_n)\in A$ is a joint action profile. Moreover, $u:=(u_1,\ldots,u_n)$, where $u_i:A_i\to\mathbb{R}$ is a real-valued utility (or payoff) function for player $i$. Also, a mixed strategy/policy for player $i$ is a probability distribution over its action set $A_i$, denoted by $\pi_i\in\Delta(A_i)$, and $\pi_i(a_i)$ denotes the probability for player $i$ to commit an action $a_i\in A_i$. The expected utility $u_i(\pi_i,\pi_{-i})$ of player $i$ can be expressed as $\mathbb{E}_{a\sim \pi}(u_i(a))$, where $\pi:=(\pi_1,\ldots,\pi_n)$ is the joint (mixed) action profile and $\pi_{-i}$ denotes the joint action profile of all players except player $i$. Similarly, let $a_{-i}$ be the joint (pure) action profile of all players except player $i$, and denote $u_i$ by $u_i(a_i,a_{-i})$ for manifesting the dependency of a joint pure action profile. The social welfare is defined as $SW(a):=\sum_{i=1}^n u_i(a)$ for a pure action profile $a\in A$, whose mixed strategy correspondence is given as $SW(\pi):=\mathbb{E}_{a\sim \pi}SW(a)$. In addition, the game is called {\em constant-sum} if for any action profile $a\in A$, it holds that $\sum_{i=1}^n u_i(a)=c_s$ for a constant $c_s$, and called {\em zero-sum} if $c_s=0$, as an illustration in Fig. \ref{ZSG}.

Note that for the case with continuous action sets, generally assumed closed and convex, they are usually called {\em continuous games}.

In what follows, the extensive-form games with imperfect information are introduced, which can reduce to ones with perfect information when information sets of each player is a singleton \cite{osborne1994course}.

{\bf Imperfect-Information Extensive-Form Games (II-EFGs).} An {\em II-EFG} is a tuple $(N,H,Z,A,P,\mu,u,I)$, where $N=[n]$ is a finite set of $n$ players, $H$ is a set of histories (i.e., nodes), representing the possible sequence of actions, and $Z\subseteq H$ denotes the set of terminal nodes, which have no further actions and award a value to each player. Outside of $N$, a different ``player'' exists, denoted $c$, representing chance decisions. Moreover, the empty sequence $\emptyset$ is included in $H$, standing for a unique root node. At a nonterminal node $h\in H$, $A(h):=\{a:(h,a)\in H\}$ is the action function assigning a set of available actions at $h$ (here $A(h)$ is different from $A$ in normal-form games, which should be clear from the context), and $P(h)$ is the player function assigning a player to the node $h$ who takes an action at that node with $P(h)=c$ if chance determines the action at $h$. And $h\sqsubset h'$ means that $h$ is led to $h'$ by a sequence of actions, i.e., $h$ is a prefix of $h'$. $u=(u_1,\ldots,u_n)$ is the set of utility functions, where $u_i:Z\to\mathbb{R}$ is the utility function of player $i$. If there is a constant $c_s$ such that $\sum_{i=1}^n u_i(z)=c_s$ for all $z\in Z$, then the game is called a {\em constant-sum game}, and a {\em zero-sum game} when $c_s=0$.

The main feature ``imperfect information'' is represented by information sets (infosets) for all players. Specifically, $I=(I_1,\ldots,I_n)$ is the set of information sets, where $I_i$ is a partition of $H_i:=\{h:P(h)=i,h\in H\}$ satisfying that $A(h_1)=A(h_2)$ and $P(h_1)=P(h_2)$ for any $h_1,h_2\in I_{i,j}$ for some $I_{i,j}\in I_i$. That is, all nodes in the same infoset of $I_i$ are indistinguishable to player $i$. Note that each node $h\in H$ is only in one infoset for each player. When all players can remember all historical information, it is called {\em perfect recall}. Formally, let $h,h',g,g'$ be histories such that $h\sqsubset h',g\sqsubset g'$, and then perfect recall means that if $g$ and $h$ do not share an infoset and each is not a prefix of the other, then $h'$ and $g'$ also do not share an infoset.

A {\em normal-form plan} (or {\em pure strategy}) of player $i$ is a tuple $a_i\in\Xi_i:=\times_{I_{i,j}\in I_i}A(I_{i,j})$, which assigns an action to each infoset of player $i$. A {\em normal-form strategy} $x_i$ means a probability distribution over $\Xi_i$, i.e., $x_i\in\Delta(\Xi_i)$. A {\em behavioral strategy} $\pi_i$ (or simply, strategy) is a probability distribution over $A(I_{i,j})$ for each infoset of player $i$. A joint strategy profile $\pi$ is composed of all players' strategies $\pi_i,i\in[n]$, i.e., $\pi=(\pi_1,\ldots,\pi_n)$, with $\pi_{-i}$ representing all the strategies except $\pi_i$. Denote by $p(I_{i,j},a)$ (or $p(h,a)$) the probability of a specific action $a$ at infoset $I_{i,j}$, and $p^{\pi}(h)$ the reach probability of history $h$ if all the players select their actions according to $\pi$. For a strategy profile, player $i$ has its total expected payoff as $u_i(\pi)=\sum_{h\in Z}p^{\pi}(h)u_i(h)$. Denote by $\Sigma_i$ the set of all possible strategies for player $i$.

A {\em best response} for player $i$ to $\pi_{-i}$ is a strategy $BR(\pi_{-i}):=\arg\max_{\pi_i'}u_i(\pi_i',\pi_{-i})$. In a two-player zero-sum game, the {\em exploitability} $e(\pi_i)$ of a strategy $\pi_i$, defined as $e(\pi_i):=u_i(\pi_i^*,\pi_{-i}^*)-u_i(\pi_i,BR(\pi_i))$, where $(\pi_i^*,\pi_{-i}^*)$ is a Nash equilibrium, as defined later. In multi-player games, the {\em total exploitability} (or {\em NashConv}) of a strategy profile $\pi$ is defined as \cite{lanctot2017unified} $e(\pi):=\sum_{i\in [n]}u_i(BR(\pi_{-i}),\pi_{-i})-u_i(\pi_i,\pi_{-i})$, and the {\em average exploitability} (or simply {\em exploitability}) is defined as $e(\pi)/|N|$, which is leveraged to measure how much can be gained by unilaterally deviating to their best response, generally interpreted as a distance from a Nash equilibrium.

Note that besides the above normal-form and extensive-form games, other classes of games may be conducive as well in adversarial games, such as Markov games (or stochastic games) \cite{littman1994markov}, where the game state changes according to a transition probability based on the current game state and players' actions, Bayesian games \cite{zamir2008bayesian}, which models game uncertainties with incomplete information, and so forth.

In what follows, some solution concepts for related games are introduced.

The Nash equilibrium is the most widely adopted notion in the literature \cite{nash1950equilibrium}.

\begin{definition}[$\epsilon$-Nash Equilibrium ($\epsilon$-NE)] \label{df1}
For both normal-form and extensive-form games, a strategy $\pi=(\pi_1^*,\ldots,\pi_n^*)$ is called an {\em $\epsilon$-NE} for a constant $\epsilon\geq 0$ if
\begin{align}
u_i(\pi_i^*,\pi_{-i}^*)\geq u_i(\pi_i,\pi_{-i}^*)-\epsilon,~~~\forall \pi_i,~ i\in N        \label{2}
\end{align}
that is, the gain is at most $\epsilon$ if any player changes its own strategy solely. Moreover, it is called an {\em NE} when $\epsilon=0$, that is, $\pi_i^*$ is a best response of $\pi_{-i}^*$ for any player $i\in [n]$, i.e., $\pi_i^*=BR(\pi_{-i}^*),\forall i\in[n]$.
\end{definition}

It is well known that there exists at least one NE in mixed strategies for games with finite number of players and finite number of pure strategies for each player \cite{nash1950equilibrium}.

Even though NE may exist for many games and it is computationally efficient for two-player zero-sum games, it is well known by complexity theory that approximating an NE in $\kappa$-player ($\kappa\geq 3$) zero-sum games and even two-player nonzero-sum games is computationally hard, that is, it is \textsf{PPAD}-complete for general games \cite{chen2009settling,daskalakis2009complexity,rubinstein2019hardness}. As an alternative, (coarse) correlated equilibrium is often considered for normal-form games in the literature, which is efficiently computable in all normal-form games, as defined in the following \cite{aumann1974subjectivity}.

\begin{definition}[$\epsilon$-Correlated Equilibrium ($\epsilon$-CE)]\label{df2}
For a normal-form game $(N,A,u)$, an {\em $\epsilon$-CE} is a probability distribution $\mu$ over $\times_{i\in [n]}A_i$ if for each player $i\in[n]$ and any swap function $\phi_i:A_i\to A_i$ (usually called strategy modification),
\begin{align}
\mathbb{E}_{a\sim \mu}[u_i(a_i,a_{-i})]\geq\mathbb{E}_{a\sim \mu}[u_i(\phi_i(a_i),a_{-i})]-\epsilon.       \label{3}
\end{align}
\end{definition}
That is, no player can gain more payoff by unilaterally deviating its action privately informed by a coordinator who samples a joint action $a=(a_1,\ldots,a_n)$ from that distribution. Furthermore, another relevant notion is defined below \cite{hannan1957approximation}.

\begin{definition}[$\epsilon$-Coarse Correlated Equilibrium ($\epsilon$-CCE)]\label{df3}
For a normal-form game $(N,A,u)$, an {\em $\epsilon$-CCE} is a probability distribution $\mu$ over $\times_{i\in [n]}A_i$ if for each player $i\in[n]$ and all actions $a_i'\in A_i$,
\begin{align}
\mathbb{E}_{a\sim \mu}[u_i(a_i,a_{-i})]\geq\mathbb{E}_{a\sim \mu}[u_i(a_i',a_{-i})]-\epsilon.       \label{4}
\end{align}
\end{definition}
The above condition looks like almost the same as that for $\epsilon$-CE, except the removal of the conditioning on the action $a_i$, by arbitrarily selecting an action $a_i'$ on their own, instead of following the action $a_i$ advised by the coordinator. For NE, CE, and CCE, it is known that they are payoff equivalent to each other in two-player zero-sum games by the minimax theorem \cite{neumann1928theorie}. Recently, the notions of CE and CCE have been extended to extensive-form games in \cite{farina2020coarse,celli2019computing}, which however have been less studied by now.

In an II-EFG, let us consider the case where all the players in ${\tt T}:=\{1,\ldots,n-1\}$ are cooperative, thus forming a team, who take actions independently and play against an adversary $n$, and $u_i=u_j,\forall i,j\in T$ and $u_n=-u_{\tt T}=-\sum_{i\in {\tt T}}u_i$, called a zero-sum single-team single-adversary extensive-form team game (or simply zero-sum {\em team game} (TG)) \cite{celli2018computational}. Before introducing the notion of team-maxmin equilibrium, it is necessary to first prepare some essentials. Let $S_i$ denote the set of action sequences of player $i$, where an action sequence of player $i$, defined by a node $h\in H$, is the ordered set of actions of player $i$ that are on the path from the root to $h$. Let $\emptyset$ be the dummy sequence to the root. A {\em realization plan} $r_i:S_i\to[0,1]$ is a function mapping each action sequence to a probability, satisfying
\begin{align}
r_i(\emptyset)&=1,       \nonumber\\
\sum_{a\in A(I_{i,j})}r_i(s_i,a)&=r_i(s_i),~~~\forall I_{i,j}\in I_i,~s_i=seq_i(I_{i,j}),    \nonumber\\
r_i(s_i')&\geq 0,~~~\forall s_i'\in S_i,           \label{1}
\end{align}
where $seq_i(I_{i,j})$ denotes the action sequence leading to $I_{i,j}$.

With the above preparations, the team-maxmin equilibrium, first introduced in \cite{von1997team}, is defined as follows \cite{celli2018computational}.

\begin{definition}[Team-Maxmin Equilibrium (TME)]\label{df4}
A {\em TME} is defined as
\begin{align}
\arg\max_{r_1,\ldots,r_{n-1}}\min_{r_n}\sum_{s=\times_{i\in N}s_i,s\in S}U_{\tt T}(s)\prod_{i=1}^n r_i(s_i),     \label{5}
\end{align}
where $U_{\tt T}$ stands for the team's utility defined by $U_{\tt T}(s):=\sum_{l\in Z'}u_{\tt T}(l)c(l)$ if at least one terminal node is achieved by the joint plan $s$ (i.e., $Z'\subseteq Z$ is nonempty) with the chance $c(l)$ determined by chance nodes, and $U_{\tt T}(s)=0$ otherwise.
\end{definition}
A TME is generally unique and it is an NE which maximizes the team's utility. In addition, the concept of $\epsilon$-TME can be similarly defined, at which both the team and the adversary can gain at most $\epsilon$ if any player unilaterally changes its strategy.

Besides the aforementioned optimal strategy concepts, it is worth noting that there are other notions as well, such as subgame perfect NE \cite{fudenberg1991game} and $\alpha$-rank \cite{omidshafiei2019alpha}, which however are beyond this survey.

\subsection{Stackelberg Games}\label{subs2.2}

Stackelberg games (SGs, or leader-follower games) can date back to Stackelberg competition introduced in \cite{von1934marktform} to model a strategic game between two firms, the leader and the follower, where the leader can take actions first. SGs, as games with sequential actions and asymmetric information, have many practical applications, for example, PROTECT, a system that the United States Coast Guard utilizes to assign patrols in Boston, New York, and Los Angeles \cite{an2013deployed}, and ARMOR, an assistant deployed in Los Angeles International Airport in 2007 for randomly scheduling checkpoints on the roadways entering the airport. In what follows, general Stackelberg games and Stackelberg security games \cite{casorran2019study} are introduced, where the second one is an important special case of general SGs.

\begin{figure}[h]
\centering
\includegraphics[width=1.6in]{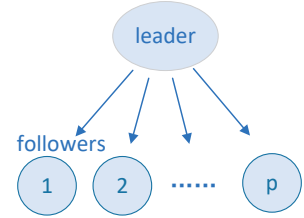}
\caption{A schematic illustration of general Stackelberg games, where directed edges mean that the leader commits an action first and then the followers play actions in response to the leader's action.}
\label{SG}
\end{figure}

{\bf General Stackelberg Game (GSG).} A GSG consists of a leader, who commits an action first, and $p$ followers, who can observe and learn the leader's strategy and then take actions in response to the leader's strategy, see Fig. \ref{SG}. Denote by $F$, $A_l$, $A_f$ the sets of $p$ followers, the leader's pure strategies, and each follower's pure strategies, respectively. The leader knows the probability of facing follower $k\in F$, denoted as $\varpi^k\in[0,1]$. Denote by $x\in\Delta(A_l)$ the mixed strategy of the leader, where the $i$-th component $x_i$ represents the probability of choosing the $i$-th pure strategy by the leader. Let $q_j^k\in\{0,1\}$ denote the decision of follower $k\in F$ to take a pure strategy $j\in A_f$ such that $\sum_{j\in A_f}q_j^k=1$ for all $k\in F$. Note that it is enough for the rational followers to only consider pure strategies \cite{conitzer2006computing}. For the leader and each follower $k\in F$, the utilities (or payoffs, rewards) of the leader and the follower are captured by a pair of matrices $(R^k,C^k)$, where $R^k\in\mathbb{R}^{|A_l|\times |A_f|}$ is the utility matrix of the leader when facing follower $k$, and $C^k\in\mathbb{R}^{|A_l|\times |A_f|}$ is the utility matrix of follower $k\in F$. Then, the expected utilities of the leader and follower $k$ can be, respectively, given as
\begin{align}
U_l(x,q)&=\sum_{i\in A_l}\sum_{j\in A_f}\sum_{k\in F}\varpi^k x_i q_j^k R_{ij}^k,       \label{6}\\
U_f^k(x,q^k)&=\sum_{i\in A_l}\sum_{j\in A_f} x_i q_j^k C_{ij}^k,                       \label{7}
\end{align}
where $q:=(q^1,\ldots,q^{|F|})$ and $q^k:=(q_1^k,\ldots,q_{|A_f|}^k)$ for each $k\in F$.

{\bf Stackelberg Security Game (SSG).} In SSG, as a specific case of GSG, the leader and followers are viewed as the defender and attackers, where the defender aims to schedule a limited number of $m$ security resources to protect (or cover) a subset of $n$ targets from the attackers' attacks, with $m<n$. The notations $F,A_l,A_f,\varpi^k,x,q_j^k$ are the same defined as in the above GSG. Noting that $|A_f|=n$ in this case, the leader's pure strategy set $A_l$ is now composed of all possible subsets of at most $m$ targets that can be safeguarded simultaneously, and $q_j^k\in\{0,1\}$ indicates whether attacker $k\in F$ attacks target $j\in[n]$. Let $c_j\in[0,1]$ be the probability of coverage of target $j\in [n]$ such that $c_j=\sum_{i\in A_l,j\in i}x_i$, where $j\in i$ connotes that the target $j$ is covered by pure strategy $i$. When facing attacker $k\in F$ who attacks target $j\in[n]$, the defender's utility is $D_c^k(j)$ if the target is covered or protected, or $D_u^k(j)$ if the target is uncovered or unprotected. The utility of attacker $k\in F$ is $A_c^k(j)$ when attacking target $j$ that is covered, or $A_u^k(j)$ when attacking target $j$ that is uncovered. It is generally assumed that $D_c^k(j)\geq D_u^k(j)$ and $A_u^k(j)\geq A_c^k(j)$, which are in line with the common sense. The expected utilities for the defender and attacker $k\in F$ is, respectively, expressed as
\begin{align}
U_d(x,q)&=\sum_{j\in A_f}\sum_{k\in F}\varpi^k q_j^k [c_jD_c^k(j)+(1-c_j)D_u^k(j)],         \label{8}\\
U_a^k(x,q^k)&=\sum_{j\in A_f} q_j^k [c_j A_c^k(j)+(1-c_j)A_u^k(j)].                       \label{9}
\end{align}

A most widely adopted solution for GSG and SSG is the so-called strong Stackelberg equilibrium, which always exists in all Stackelberg games \cite{leitmann1978generalized,casorran2019study}. Recall that it is enough for each follower to play pure strategies.

\begin{definition}[Strong Stackelberg Equilibrium (SSE)]\label{df5}
A strategy profile $(x^*,\{q^{k*}\}_{k\in F})$ for a GSG forms an {\em SSE}, if
\begin{enumerate}
  \item $x^*$ is optimal for the leader:
  \begin{align}
  (x^*)^\top R^k q^{k*}\geq x^\top R^k \mathcal{R}^k(x),~\forall x\in\Delta(A_l),~\forall k\in F    \nonumber
  \end{align}
  where $\mathcal{R}^k(x)$ denotes the attacker $k$'s best response against $x$.
  \item Each follower $k$ always plays a best-response, i.e.,
  \begin{align}
  (x^*)^\top C^k q^{k*}\geq (x^*)^\top C^k q^k,~~\forall~\text{feasible}~q^k.    \nonumber
  \end{align}
  \item Each follower $k$ breaks ties in favor of the leader:
  \begin{align}
  (x^*)^\top R^k q^{k*}\geq (x^*)^\top R^k \mathcal{R}^k(x^*),~~\forall k\in F.      \nonumber
  \end{align}
\end{enumerate}
\end{definition}
The tie-breaking rule is reasonable in cases of indifference since the leader can often induce the favorable equilibrium by choosing a strategy arbitrarily close to the equilibrium that makes the follower prefer the desired strategy \cite{von2010market}. When the tie-breaking rule is in favor of the followers, then the equilibrium is called weak Stackelberg equilibrium (WSE), which however does not always exist \cite{bacsar1998dynamic}. Moreover, the concept of SSE can be similarly defined for SSGs.

\subsection{Zero-Sum Differential Games}\label{subs2.3}

Differential games (DGs), also known as dynamic games \cite{bacsar1998dynamic}, are a natural extension of sequential games to continuous-time scenario, which are expressed by differential equations and first introduced by Isaacs \cite{issacs1965diff}. DGs can be regarded as an extension of optimal control \cite{lewis2012optimal}, which usually has a single decision maker with a single objective function, while multiple players are involved in a DG with noncooperative objectives. Since this survey is concerned with adversarial games, zero-sum DGs (mostly involving two players in the literature) are considered here, although many other types of DGs emerge in the literature, including nonzero-sum differential games, mean-field games, differential graphical games, Dynkin games, and so on \cite{buckdahn2011some,friedman2013differential}.

\begin{figure}[h]
\centering
\includegraphics[width=2.2in]{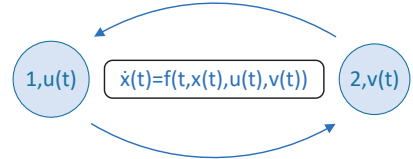}
\caption{A schematic illustration of two-player differential games.}
\label{DG}
\end{figure}

A {\em two-player zero-sum differential game (TP-ZS-DG)} is described by a dynamical system as
\begin{align}
\dot{x}(t)&=f(t,x(t),u(t),v(t)),~~t\in[t_0,T]      \nonumber\\
x(t_0)&=x_0,~~u(t)\in U,~~v(t)\in V,                  \label{10}
\end{align}
where $x(t)\in\mathbb{R}^d$ is the state vector at time $t$, $t_0$ is the initial time, $x_0$ is the initial state, $U\subseteq\mathbb{R}^{m_1}$, $V\subseteq\mathbb{R}^{m_2}$ are control constraints for players $1$ and $2$, respectively, $u(t)$ and $v(t)$ are control actions (or signals) for player $1$ and $2$, respectively, and $f:[t_0,T]\times \mathbb{R}^d\times U\times V\to\mathbb{R}^d$ is the dynamics, as illustrated in Fig. \ref{DG}.

For different setups in the literature, distinct cost functions are generally employed, most of which, however, are either based on or variants of an essential and important cost function, as given below:
\begin{align}
J(u(\cdot),v(\cdot))=\int_{t_0}^T f_0(t,x(t),u(t),v(t))dt+\phi(x(T)),     \label{11}
\end{align}
where $f_0:[t_0,T]\times \mathbb{R}^d\times U\times V\to\mathbb{R}$ is the {\em running cost} (or {\em stage cost}) and $\phi:\mathbb{R}^d\to\mathbb{R}$ is the {\em terminal cost} (or {\em final cost}).

With (\ref{11}), the goal of DG (\ref{10}) is for player $1$ to minimize the cost $J$, while player $2$ aims at maximizing it, i.e.,
\begin{align}
\min_{u(\cdot)\in U}\max_{v(\cdot)\in V} J(u(\cdot),v(\cdot)).        \label{12}
\end{align}
For (\ref{12}), the optimal cost of $J$ is called the {\em value} of the game, expressed as a {\em value function} $\psi(t,x)$. Moreover, the solution notion is still the NE as in normal-form and extensive-form zero-sum games, also called {\em minimax equilibrium} (or {\em minimax point}, {\em saddle point}) in the literature since the studied problem is in fact a saddle point game (or saddle point problem/optimization).

Note that dynamics (\ref{10}) is deterministic. In the meantime, stochastic DGs have also been addressed in the literature, described by stochastic differential equations with the standard Brownian motion \cite{buckdahn2011some}. It is also noteworthy that the above DGs are usually studied under a set of assumptions, such as the compactness of $U,V$, and the Lipschitz continuity of $f,f_0,\phi$, among others \cite{friedman2013differential}.

Finally, the main features of the aforementioned games are summarized in Table \ref{tb_models}.

\begin{table*}[!t]
\renewcommand{\arraystretch}{1.3}
\caption{{\upshape Main features of various adversarial games.}}
\label{tb_models}
\centering
\begin{threeparttable}
\begin{tabular}{c|c|c|c|c}
\hline
 Game models & Player numbers & Action order & Information & Dynamics \\
\hline\hline
Zero-sum NFG & $\geq 2$ & mostly simultaneous & symmetric & \makecell[c]{discrete-time\\continuous-time} \\
\hline
Zero-sum EFG & \makecell[c]{$\geq 2$ (perfect information)\\mostly $2$ (imperfect information)} & sequential & symmetric & mostly discrete-time \\
\hline
GSG and SSG & \makecell[c]{mostly one-leader \\$p$-follower} & sequential & asymmetric & mostly discrete-time \\
\hline
Zero-sum DG & mostly $2$ & mostly simultaneous & mostly symmetric & continuous-time \\
\hline
\end{tabular}
\end{threeparttable}
\end{table*}

\section{Research Classification and Frontiers}\label{sec3}

This section aims to succinctly summarize the relevant literature for zero-sum games, GSGs, SSGs, and TP-ZS-DGs along with the emerging state-of-the-art research. However, the relevant literature on adversarial games is too vast to cover it all, and thus only the literature of our interest is reviewed throughout this survey.

\subsection{Zero-Sum Games (ZSGs)}\label{sec3.1}

Both normal-form and extensive-form ZSGs studied in the literature can be generally categorized into the following main aspects: bilinear games, saddle point problems, multi-player ZSGs, team games, and imperfect-information ZSGs, as discussed below.
\begin{enumerate}
  \item {\em Bilinear Games.} Bilinear games are simple models for delineating two-player games, generally in normal-form as \cite{garg2011bilinear}: maximizing utilities $x^\top Ay$ and $x^\top By$ for players $1$ and $2$, respectively, where $A\in\mathbb{R}^{m\times n}$ and $B\in\mathbb{R}^{m\times n}$ are payoff matrices, subject to strategy sets $x\in X:=\{x\in\mathbb{R}^{m}:R_1 x=r_1,x\in\mathbb{R}_+^m\}$ and $y\in Y:=\{y\in\mathbb{R}^n:R_2y=r_2,y\in\mathbb{R}_+^n\}$ with some $R_1\in\mathbb{R}^{k_1\times m},R_2\in\mathbb{R}^{k_2\times n}$ and $r_1\in\mathbb{R}^{k_1},r_2\in\mathbb{R}^{k_2}$. A bilinear game is usually denoted by the payoff matrices pair $(A,B)$, which is zero-sum when $B=-A$, and as an important notion, the rank of a game $(A,B)$ is defined as the rank of matrix $A+B$. Several interesting games can be viewed as special cases of bilinear games, such as bimatrix games \cite{lemke1964equilibrium,anagnostides2020solving,dinh2021online}, where $R_1={\bf 1}_m^\top,R_2={\bf 1}_n^\top$ and $r_1=r_2=1$, imitation games (a special case of bimatrix games with $B=I$) \cite{murhekar2020approximate}, and the Colonel Blotto game (i.e., two colonels simultaneously allocate their troops across different battlefields) \cite{borel1921theorie}. In addition, multi-player polymatrix games \cite{howson1972equilibria} can also be equivalently transformed to bilinear games \cite{garg2011bilinear}. Generally speaking, the existing literature mainly focuses on the computational complexity and polynomial-time algorithm design for approximating NE of bilinear games \cite{sengodan2020linear}, bimatrix games \cite{deligkas2022polynomial}, polymatrix games \cite{deligkas2020tree}, and the Colonel Blotto game \cite{seddighin2019campaigning}. Recently, it is shown that NE computation in two-player nonzero-sum games with rank $\geq 2$ is {\sf PPAD}-hard \cite{mehta2018constant,boodaghians2020smoothed}. And computing a $1/n_s^c$-approximate NE is {\sf PPAD}-hard even for imitation games for any $c>0$ \cite{murhekar2020approximate}, where $n_s$ is the number of moves available to the players, and a polynomial-time algorithm was developed for finding an approximate NE in \cite{murhekar2020approximate}. Also, computing an NE in a tree polymatrix game with twenty actions per player is {\sf PPAD}-hard \cite{deligkas2020tree}, and a polynomial-time algorithm for $1/3$-approximate NE in bimatrix games was proposed in \cite{deligkas2022polynomial}, which is the state-of-the-art in the literature. For the Colonel Blotto game, efficient and simple algorithms have been recently provided in \cite{behnezhad2019optimal,behnezhad2022fast,beaglehole2022efficient}, and meanwhile, various scenarios have been extended for this game, including dynamic Colonel Blotto game \cite{leon2021bandit}, generalized Colonel Blotto and generalized lottery Blotto games \cite{vu2019approximate}, and multi-player cases \cite{boix2021multiplayer,beaglehole2022efficient}. Furthermore, bilinear games are generalized to hidden bilinear games in \cite{vlatakis2019poincare}, where the inputs controlled by players are first processed by a smooth function, i.e., a hidden layer, before coming into the conventional bilinear games.
  \item {\em Saddle Point Problems (SPPs).} SPPs are also called saddle point optimization, min-max/minimax games, or min-max/minimax optimization in the literature. The formulation of a {\em general SPP} \cite{zhang2022near} is given as $\min_{x\in X}\max_{y\in Y}f(x,y)$, where $X\subseteq \mathbb{R}^{m}$ and $Y\subseteq\mathbb{R}^n$ are closed and convex, possibly the entire Euclidean spaces or their compact subsets. For general SPPs, besides zero-sum bilinear games, other two types have been extensively considered, that is, non-bilinear SPPs and bilinear SPPs. A {\em non-bilinear SSP} \cite{hamedani2021primal,tominin2021accelerated} is expressed as $\min_{x\in X}\max_{y\in Y} f(x)+\Theta(x,y)-h(y)$, where $\Theta$ is a general coupling function, and as a special case, when $\Theta(x,y)=x^\top C y$ with $C\in\mathbb{R}^{m\times n}$, the game is called a {\em bilinear SPP} \cite{xie2021dippa,kovalev2021accelerated,thekumparampil2022lifted} due to the bilinear coupling. The existing research mainly centers on equilibrium existence, computational and sampling complexity, and efficient algorithm design, for instance, as done in the aforementioned recent works. Meanwhile, various scenarios have been investigated in the literature, including projection-free methods by applying the Frank-Wolfe algorithm \cite{gidel2017frank,chen2020efficient}, nonconvex-nonconcave general SPPs \cite{li2021complexity,hsieh2021limits}, linear last-iterate convergence \cite{wei2020linear}, SSPs with adversarial bandits and delays \cite{bistritz2021no}, periodic zero-sum bimatrix games with continuous strategy spaces \cite{fiez2021online}, compositional SSPs \cite{gao2021convergence}, decentralized setup \cite{beznosikov2021distributed}, and hidden general SPPs \cite{vlatakis2021solving}, where the controlled inputs are first fed into smooth functions whose outputs are then treated as inputs for the traditional general SPPs. Finally, it is noteworthy that the general SPPs with sequential actions have also been studied, called {\em min-max Stackelberg games}, for example, the recent work \cite{goktas2021convex} with dependent feasible sets.
  \item {\em Multi-Player Zero-Sum Games (MP-ZSGs).} The above discussed games usually involve two players. It is well known that approximating an NE in multi-player zero-sum games and even two-player nonzero-sum games is {\sf PPAD}-complete \cite{chen2009settling,daskalakis2009complexity,rubinstein2019hardness}. Moreover, it is known that multi-player symmetric zero-sum games might have only asymmetric equilibria, which is consistent with that of two-player and multi-player symmetric nonzero-sum games, but in contrast with the case in two-player symmetric zero-sum games that always have symmetric equilibria (if equilibria exist) \cite{xefteris2015symmetric}. In the literature, most of works focus on multi-player zero-sum polymatrix games (also called network matrix games in some works), where the utility of each player is composed of the sum of utilities gained by playing with its neighbors in an undirected graph \cite{howson1972equilibria}. The authors in \cite{cai2011minmax} generalized von Neumann's minimax theorem to multi-player zero-sum polymatrix games, thus, implying convexity of equilibria, polynomial-time tractability, and convergence of no-regret learning algorithms to NEs, and last-iterate convergence was studied in \cite{anagnostides2022last} for multi-player polymatrix zero-sum games. $O(1/T)$ time-average convergence was established by using alternating gradient descent in \cite{bailey2021left}, where $T$ is the time horizon. Moreover, it is shown that for continuous-time algorithms, time-average convergence may fail even in a simple periodic multi-player zero-sum polymatrix game or replicator dynamics, but being Poincar\'{e} recurrent in \cite{fiezonline,skoulakis2021evolutionary}. What's more, it is realized that mutual cooperations among players may benefit more than pursuing selfish exploitability, and in this case, team/alliance formation is also studied in the literature, for example, \cite{hughes2020learning}, where it was demonstrated that team formation may be seen as a social dilemma. Additionally, other pertinent research encompasses multi-player general-sum games \cite{ganzfried2020fast,anagnostides2021near,anagnostides2022faster} and machine learning based studies \cite{gidel2020multi}, etc.
  \item {\em Team Games (TGs).} Generically, team games refer to those games where at least one team exists with the cooperation of team members with communications either before the play, or during the play, or simultaneously before and during the play, or without any communications. In general, team games in the literature can be classified from two perspectives. One perspective depends upon the team number, i.e., one-team games (or adversarial team games) \cite{zhang2020converging}, where players in the team enjoying the same utility function play against an adversary independently, and two-team games \cite{kalogiannis2021teamwork} consisting of two teams in a game. The other perspective is on perfect-information and imperfect-information games. For team games, TME is an important solution concept, for which it is known that computing a TME is {\sf FNP}-hard and inapproximable in additive sense \cite{hansen2008approximability,borgs2010myth}. Even though, efficient algorithms for computing a TME in perfect-information zero-sum NFGs have been developed until now, e.g., \cite{zhang2020converging}. Meanwhile, a class of zero-sum two-team games in perfect-information normal-form was studied in \cite{kalogiannis2021teamwork}, where finding an NE is shown to be {\sf CLS}-hard, i.e., unlikely to have a polynomial-time NE computing algorithm. Moreover, as two-team games, two-network zero-sum games are also addressed, where each network is thought of as a team \cite{gharesifard2013distributed,lou2015nash,huang2021no}. For imperfect-information zero-sum team games, the researchers have investigated a variety of scenarios centering around the computational complexity and efficient algorithms, such as one-team games \cite{celli2018computational,zhang2020computing,carminati2022public}, one-team game with two members in the team \cite{farina2020faster}, the computation of team correlated equilibrium in two-team games \cite{zhang2021team}.
  \item {\em Imperfect-Information ZSGs (II-ZSGs).} Unlike perfect-information games, such as Chess, Go and backgammon, II-ZSGs, involving individual players' primate information that is hidden to other players, are more challenging due to information privacy and uncertainty, especially for large games with large action spaces and/or infosets. For example, the game of heads-up (i.e., two-player) limit Texas Hold'em poker, with over $10^{14}$ infosets, has been a challenging problem for AI for over $10$ years, before essentially solved by Cepheus \cite{tammelin2015solving}, the first computer program for handling imperfect information games that are played competitively by humans. Also, the game of no-limit Texas Hold'em poker has more than $10^{161}$ infosets, for which DeepStack \cite{moravvcik2017deepstack} and Libratus \cite{brown2018superhuman} are the first line of AI agents/algorithms to defeat professional humans in heads-up no-limit Texas Hold'em poker. As such, most of research focuses on the computing of NEs in two-player II-ZSGs in the literature \cite{munos2020fast,farina2021better}, aiming to develop efficient superhuman AI agents in face of the challenges of imperfect information, large models and uncertainties. To handle large games with imperfect information, several techniques have been successively proposed, for exmaple, pruning, abstraction, and search \cite{brown2017safe,brown2018depth,brown2020equilibrium}. Roughly speaking, pruning aims to avoid traversing the whole game tree for an algorithm while simultaneously ensuring the same convergence, including regret-based pruning, dynamic thresholding, best-response pruning, and so on \cite{marsland1986review}. Abstraction aims to generate a smaller version of the original game by bucketing similar infosets or actions, while maintaining as much as possible the strategic features of the original game \cite{sandholm2015solving}, mainly including information abstraction and action abstraction. Meanwhile, search tries to improve upon the (approximate) solution of a game abstraction, which may be far from the true solution of the original game, by seeking a more precise equilibrium solution for a faced subgame, such as depth-limited search \cite{brown2018depth,schmid2021search}. Moreover, it has been shown recently that some two-player poker games can be represented as perfect-information sequential Bayesian extensive games with efficient implementation \cite{kovavrik2021fast}. The authors in \cite{farina2022kernelized} recently bridged several standing gaps between NFG and EFG learning by directly transferring desirable properties in NFGs to EFGs, guaranteeing simultaneously last-iterate convergence, lower dependence on the game size, and constant regret in games. Besides, bandit feedback is of practical importance in real-world applications for II-ZSGs \cite{meng2022generalized,bai2022near}, where only the interactive trajectory and the payoff of the reached terminal node can be observed without prior knowledge of the game, such as the tree structure, the observation/state space, and transition probabilities (for Markov games) \cite{kozuno2021model}. On the other hand, multi-player II-ZSGs are more challenging and thus have been less researched except for a handful of works, for example, Pluribus \cite{brown2019superhuman}, the first multi-player poker agent, has defeated top humans in six-player no-limit Texas Hold'em poker (the most prevalent poker in the world) \cite{blair2019ai}, and other endeavors \cite{wu2019hierarchical,tian2020joint,ganzfried2020parallel,bai2022near,weilin2021imperfect}. Aside from deterministic methods, AI approaches have achieved great success in II-SSGs based on reinforcement learning, deep neural networks and so on \cite{heinrich2016deep,moravvcik2017deepstack,li2019double,farnia2020gans,gruslys2020advantage,ye2020towards,ye2020mastering,kozuno2021model,
schmid2021player,phillips2021reinforcement,fu2021actor,wang2022unified,feng2021neural,feng2021discovering}, for instance, AlphaGo (the first AI agent to achieve superhuman level in Go) \cite{silver2016mastering}, AlphaZero (with initial training independent of human data and Go-specific features, reaching state-of-the-art performance in Go, Chess and Shogi with minimal domain knowledge) \cite{silver2017mastering}, and DeepStack \cite{moravvcik2017deepstack}, to name a few. More details can refer to a recent survey for AI in games \cite{yin2021ai}. Note that other closely related research subsumes imperfect-information general-sum games with full and bandit feedback \cite{celli2019learning,celli2020no,song2022sample}, two-player zero-sum Markov games \cite{wei2021last}, and multi-player general-sum Markov games \cite{mao2022provably}.
\end{enumerate}

It should be noted that incomplete information is also important in adversarial games, mainly comprising of Bayesian games (cf. a recent survey \cite{insua2020advances}).

\subsection{Stackelberg Games}\label{sec3.2}

Stackelberg games are roughly summarized from four perspectives, i.e., GSGs, SSGs, continuous Stackelberg games, and incomplete-information Stackelberg games.

\begin{enumerate}
  \item {\em GSGs.} The research on GSGs mainly lies in three aspects, i.e., computational complexity, solution methods, and their applications. For computational complexity, when only having one follower in GSGs, it is known that the problem can be solved in polynomial time, while it is {\sf NP}-hard for the multiple followers case \cite{conitzer2006computing}. Regarding solution methods, there are an array of proposed methods in the literature, but primarily depending upon approaches for coping with linear programming (LP) and mixed integer linear programming (MILP), including cutting plane methods, enumerative methods, and hybrid methods, among others \cite{casorran2017formulations,arriagada2021benders}. Note that both GSGs and SSGs can be formulated as bilevel optimization problems \cite{casorran2017formulations,arriagada2021benders}, where bilevel optimization has a hierarchical structure with two level optimizations, one lower level optimization (follower) nested in another upper level optimization (leader) as constraints, which is an active research area unto itself \cite{dempe2018bilevel}. As for practical applications, a multitude of real-world problems have been tackled using Stackelberg games, such as economics \cite{li2017review}, smart grid \cite{maharjan2013dependable,yu2015real}, wireless networks \cite{yang2013coping}, dynamic inspection problems \cite{guzman2021sequential}, industrial internet of things \cite{jiang2021iiot}, etc. It should be noted that other relevant cases have also been studied in the literature, such as multi-leader cases \cite{leyffer2010solving,zhang2016multi,mallozzi2017multi,tran2020resource,castiglioni2021committing}, the case with bounded rationality \cite{pita2010robust}, and general-sum games \cite{bai2021sample}, etc.
  \item {\em SSGs.} In general, SSGs can be classified by the functionality of security resources. To be specific, when every resource is capable of protecting every target, it is called {\em homogeneous} resources, and when resources are restricted to protecting only some subset of targets, it is called {\em heterogeneous} resources. Meanwhile, resources can also be distinguished by how many targets they are able to cover simultaneously, and in this case, a notion, called {\em schedule}, is assigned to a resource with the {\em size} of the schedule being defined to be the number of targets that can be simultaneously covered by the resource, including the case with size $1$ \cite{kiekintveld2009computing} and greater than $1$ \cite{jain2010software}. For these scenarios, the computational complexity was addressed in \cite{korzhyk2010complexity} when existing a single attacker, as shown in Table \ref{tbsg1}. With regard to solution methods, similar methods for solving GSGs can be applied to handle SSGs. Moreover, the practical applications of SSGs encompass wildlife protection \cite{fang2016green}, passenger screening at airports \cite{brown2016one}, crime prevention \cite{zhang2016keeping}, cyber-security \cite{dasgupta2021adversary}, information security \cite{galinkin2021information}, border patrol \cite{bucarey2017building,bucarey2021coordinating}, and so forth. In the meantime, there are other scenarios addressed in the literature, like multi-defender cases \cite{lou2015equilibrium,mutzari2022robust}, Bayesian generalizations \cite{li2016catcher}, and the case with bounded rationality \cite{wang2019repeated} and ambiguous information \cite{ma2021decision}, etc.
\begin{table}[!ht]
\renewcommand{\arraystretch}{1.3}
\caption{{\upshape Complexity results with a single attacker \cite{korzhyk2010complexity}.}}
\label{tbsg1}
\centering
\begin{threeparttable}
\begin{tabular}{c|c|c|c}
\hline
\multirow{2}*{SSGs} & \multicolumn{3}{c}{Size of schedule} \\
\cline{2-4}
~ & $1$ & $2$ & $\geq 3$ \\
\hline\hline
Homogeneous resources & {\sf P} & {\sf P} & {\sf NP}-hard \\
\hline
Heterogeneous resources & {\sf P} & {\sf NP}-hard & {\sf NP}-hard \\
\hline
\end{tabular}
\end{threeparttable}
\end{table}
  \item {\em Continuous Stackelberg games.} This sort of games mean Stackelberg games with continuous strategy spaces. In general, there exist two players, a leader and a follower, who have cost functions $f_1:\Omega\to\mathbb{R}$ and $f_2:\Omega\to\mathbb{R}$ with $\Omega:=X\times Y$, respectively, where $X\subseteq\mathbb{R}^{d_1}$ and $Y\subseteq\mathbb{R}^{d_2}$ are closed convex and possibly compact strategy sets for the leader and the follower, respectively. Then, the problem can be formally written as
      \begin{align}
      \min_{x\in X}\{f_1(x,y): y\in\arg\min_{y\in Y}f_2(x,y)\},          \label{ssg1}
      \end{align}
      where the follower still takes actions in response to the leader after the leader makes its decision first. In this case, a strategy $x^*\in X$ of the leader is called a {\em Stackelberg equilibrium} strategy \cite{fiez2019convergence} if
      \begin{align}
      \sup_{y\in BR(x^*)}f_1(x^*,y)\leq \sup_{y\in BR(x)}f_1(x,y),~\forall x\in X_1     \label{ssg2}
      \end{align}
      where $BR(x)=\{y\in Y:f_2(x,y)\leq f_2(x,y'),\forall y'\in Y\}$ is the best response of the follower against $x$. Along this line, a hierarchical Stackelberg v/s Stackelberg game was studied in \cite{kulkarni2015existence}, where the first general existence result for the games' equilibria is established without positing single-valuedness assumption on the equilibrium of the follower-level game. Furthermore, the connections between the NE and Stackelberg equilibrium were addressed in \cite{fiez2019convergence}, where convergent learning dynamics are also proposed by using Stackelberg gradient dynamics that can be regarded as a sequential variant of the conventional gradient descent algorithm, and both zero-sum and general-sum games are considered therein. Additionally, as a special case of the above game (\ref{ssg1}), {\em min-max Stackelberg games} are paid attention to as well, where the problem is of the form $\min_{x\in X}\max_{y\in Y}f(x,y)$ with $f:\Omega\to \mathbb{R}$ being the cost function. This problem has been investigated in the literature, especially for the case with dependent strategy set \cite{goktas2021convex,goktas2022robust}, i.e., inequality constraints $g(x,y)\geq {\bf 0}$ are imposed for the follower for some function $g:\Omega\to\mathbb{R}$, for which the prominent minimax theorem \cite{neumann1928theorie} does not hold any more.
  \item {\em Incomplete-Information Stackelberg Games.} Incomplete information means that the leader can only access partial information or cannot access any information of the followers' utility functions, moves, or behaviors. This is in contrast with the traditional Stackelberg games, where the followers' information is available to the leader. This weak scenario has been extensively considered in recent years motivated by practical applications. For example, the authors in \cite{maffioli2019dealing} studied situations in which only partial information on the attacker behavior can be observed by the leader. And a single-leader-multiple-followers SSG was considered in \cite{cheng2022single} with two types of misinformed information, i.e., misperception and deception, for which a stability criterion is provided for both strategic stability and cognitive stability of equilibria based on hyper NE. Additionally, one of interesting directions is information deceptions of the follower, that is, the follower is inclined to deceive the follower by sending misinformation, such as fake payoffs, to the leader in order to benefit itself as much as possible, while, at the same time, the leader needs to distinguish the deception information for minimizing its loss incurred by the deception. Recently, an interesting result on the nexus between the follower's deception and the leader's maximin utility is obtained for optimally deceiving the leader in \cite{birmpas2021optimally}, that is, through deception, almost any (fake) Stackelberg equilibrium can be induced by the follower if and only if the leader procures at least their maximin utility at this equilibrium.
\end{enumerate}


\subsection{Zero-Sum Differential Games}\label{sec3.3}

According to the existing literature, zero-sum DGs are categorized by five main dimensions, which however are not mutually exclusive, but from different angles of studied problems, i.e., linear-quadratic DGs, DGs with nonlinear dynamical systems, Stackelberg DGs, stochastic DGs, and terminal time and state constraint.

\begin{enumerate}
  \item {\em Linear-Quadratic DGs.} This relatively simple model has been widely studied for DGs, where dynamical systems are linear differential equations and cost functions are quadratic \cite{lukes1971global,engwerda2009linear}. In general, linear-quadratic DGs are analytically and numerically solvable, which can find a variety of applications in reality, such as pursuit-evasion problem \cite{shinar2008solvability,weintraub2020introduction}. Recently, singular linear-quadratic DGs were studied in \cite{gibali2021analytic}, which cannot be handled either using the Isaacs MinMax principle or the Bellman-Isaacs equation approach, and to solve this problem, an interception differential game was introduced with appropriate regularized cost functional and dual representation. The authors in \cite{huang2021defending} studied a linear-quadratic-Gaussian asset defending differential game where the state information of the attacker and the defender is not accessible to each other, but the trajectory of a moving asset is known by them. Meanwhile, a two-player linear-quadratic-Gaussian pursuit-evasion DG was investigated in \cite{huang2021pursuit} with partial information and selected observations, where the state of one player can be observed any time preferred by the other player and the cost function of each player consists of the direct cost of observing and the implicit cost of exposing his state. A linear-quadratic DG with two defenders and two attackers against a stationary target was considered in \cite{garcia2020two}. Two-player mean-field linear-quadratic stochastic DGs in an infinite horizon was investigated in \cite{li2020mean}, where the existence of both open-loop and closed-loop saddle points is studied by resorting to coupled generalized algebraic Riccati equations.
  \item {\em Nonlinear DGs.} The DGs with nonlinear state dynamics have also been taken into account in the literature, given that many practical applications cannot be dealt with by linear-quadratic DGs. For example, the authors in \cite{zhang2011iterative} considered a class of nonlinear TP-ZS-DGs by appealing to an adaptive dynamic programming. TP-ZS-DGs were addressed in \cite{song2019robust} by proposing an approximate optimal critic learning algorithm based on policy iteration of a single neural network. Nonlinear DGs were also considered with time delays \cite{lukoyanov2003functional,plaksin2019hamilton,meng2020alinear} and fractional-order systems \cite{gomoyunov2020dynamic}, and then were studied in \cite{moon2019zero} with the dynamical system depending on the system's distribution and the random initial condition. Besides two players, multi-player zero-sum DGs with uncertain nonlinear dynamics were considered and tackled using a new iterative adaptive dynamic programming algorithm in \cite{liu2014multiperson}.
  \item {\em Stackelberg DGs.} Motivated by the fact of sequential actions in some practical applications, like Stackelberg games, DGs with sequential actions, called Stackelberg DGs, have been broadly addressed in the literature. For instance, a linear-quadratic Stackelberg DG was considered in \cite{shi2020linear} with mixed deterministic and stochastic controls, where the follower can select adapted random processes as its controller. The Stackelberg DG was employed to fight terrorism in \cite{megahed2019stackelberg}. Then, the authors in \cite{lee2021hamilton} investigated two classes of state-constrained Stackelberg DGs with a nonzero running cost and state constraint, for which Hamilton-Jacobi equations are established.
  \item {\em Stochastic DGs.} In many realistic problems, the dynamics of a concerned system may not be completely modelled, but undergoing some uncertainties and/or noises, and thereby, stochastic differential equations have been leveraged to model the system dynamics in stochastic DGs \cite{elliott1981optimal,moon2018risk}. In this respect, the authors in \cite{sun2021two} considered two-person zero-sum stochastic linear-quadratic DGs, along with the investigation of the open-loop saddle point and the open-loop lower and upper values. A class of stochastic DGs with ergodic payoff were studied in \cite{li2021value}, where it is not necessary for the diffusion system to be non-degenerate. In addition, linear-quadratic stochastic Stackelberg DGs were taken into consideration in \cite{shi2017linear} with asymmetric roles for players, \cite{moon2021linear} for jump-diffusion systems, \cite{sun2021zero} without the solvability assumption on the associated Riccati equations, and \cite{huang2021robust} with model uncertainty. And a Stackelberg stochastic DG with nonlinear dynamics and asymmetric noisy observation was addressed in \cite{zheng2021stackelberg}.
  \item {\em Terminal Time and State Constraint.} A basic classification of zero-sum DGs can be made based on terminal time and state constraint, that is, whether the terminal time is finite (including two cases, i.e., a fixed constant or a variable to be specified) or infinite, and whether the system state is unconstrained or constrained. Along this line, the case with fixed terminal time and unconstrained state was first addressed \cite{evans1984differential}, and the state-constrained case with fixed terminal time was also studied \cite{altarovici2013general}. Meanwhile, the case with the terminal time being a variable was investigated in the literature, such as \cite{mitchell2005time} without state constraints and \cite{margellos2011hamilton,fisac2015reach} in presence of state constraints but with zero running-cost. Recently, the case with nonzero state constraint and underdetermined terminal time was investigated in \cite{lee2021hamilton}. Besides the above finite horizon cases, the infinite horizon case has also been considered in the literature, e.g., \cite{li2020mean,asri2021deterministic}.
\end{enumerate}

Last, it is worth pointing out that other possible forms of zero-sum DGs exist in the literature, such as the case with continuous and/or impulse controls \cite{asri2021deterministic}, mean-field DGs \cite{moon2020linear,sun2021mean}, risk-sensitive zero-sum DGs \cite{moon2018risk} and so forth.

\section{Prevailing Algorithms and Approaches}\label{sec4}

This section aims at encapsulating some main efficient algorithms and approaches for handling the reviewed adversary games as discussed in Section \ref{sec2}.

\subsection{Zero-Sum Normal- and Extensive-Form Games}\label{sec4.1}

The bundle of algorithms can be roughly divided into two parts according to their applicabilities to normal-form games or imperfect-information extensive-form games.

For normal-form games, a large number of algorithms have so far been proposed, e.g., regret matching (RM for short, first proposed by Hart and Mas-Colell in 2000 \cite{hart2000simple}), RM+ \cite{tammelin2014solving}, fictitious play \cite{brown1951iterative,ganzfried2020fictitious}, double oracle \cite{mcmahan2003planning}, online double oracle \cite{dinh2021online}, and among others. Wherein, the most prevalent algorithms are based on regret learning, usually called no-regret (or sublinear) learning algorithms, depending external and internal regrets in general, as defined below.

The {\em external regret} and {\em internal regret} \cite{xu2020distributed} for each player $i\in[n]$ are, respectively, defined as
\begin{align}
R_i^E&:=\max_{a_i\in A_i}\sum_{t=1}^T [u_i(a_i,\pi_{-i}^t)-u_i(\pi^t)],                             \label{al-zs1}\\
R_i^I&:=\max_{a_i',a_i\in A_i}\sum_{t=1}^T {\bf 1}_{a_i^t=a_i^A}[u_i(a_i',a_{-i}^t)-u_i(a^t)],       \label{al-zs2}
\end{align}
where the superscript $t$ stands for the iteration number, $T$ is the time horizon, and ${\bf 1}_E$ is the indicator function with an event $E$. Generally speaking, the external regret measures the greatest regret for not playing actions $a_i$'s, and the internal regret indicates the greatest regret for not swapping to action $a_i'$ when each time actually playing action $a_i^A$. Note that weighted external and internal regrets are also defined by adding a weight at each time $t$ \cite{zhang2022equilibrium}, and other regrets are considered as well in the literature, including swap regret \cite{anagnostides2021near} and several dynamic/static NE-based regrets \cite{lu2020online,meng2021decentralized,meng2022decentralized,zhang2022no,li2022survey}.

With regrets at hand, it is now ready to present two of most widely employed algorithms, i.e., optimistic (or predictive) follow the regularized leader (Optimistic FTRL for brevity) and optimistic mirror descent (OMD for short) \cite{anagnostides2022last}, which are, respectively, given as
\begin{align}
x^{t+1}=\arg\max_{x\in\mathcal{X}}\Big\{\alpha\Big\langle x,m^{t}+\sum_{\tau=1}^t g^t\Big\rangle-R(x)\Big\},     \label{al-zs3}
\end{align}
and
\begin{align}
x^{t+1}&=\arg\max_{x\in \mathcal{X}}\{\alpha\langle x,m^{t}\rangle-D_R(x,\hat{x}^t)\},          \nonumber\\
\hat{x}^{t+1}&=\arg\max_{\hat{x}\in\mathcal{X}}\{\alpha\langle \hat{x},g^t\rangle-D_R(\hat{x},\hat{x}^t)\},         \label{al-zs4}
\end{align}
where $\mathcal{X}$ is a generic closed convex constraint set, $\alpha>0$ is the stepsize, $g^t$ is a subgradient of a function $f^t$ returned by the environment after the player commits an action at time $t$, $m^t$ is a subgradient prediction, often assuming $m^t=g^t$ in the literature, and $R(x)$ is a strongly convex function, serving as the base function for defining the Bregman divergence $D_R(x,y):=R(x)-R(y)-\langle \nabla R(y),x-y\rangle$ for any $x,y\in\mathbb{R}^d$.

Note that many widely employed algorithms, such as optimistic gradient descent ascent (OGDA) \cite{wei2020linear} and optimistic multiplicative weights update (OMWU, or optimistic hedge) \cite{daskalakis2021near}, are special cases or variants of optimistic FTRL and OMD, and other different efficient algorithms also exist such as optimistic dual averaging (OptDA) \cite{hsieh2021adaptive}, greedy weights \cite{zhang2022equilibrium}, and so forth.

For imperfect-information games, the most popular algorithms are counterfactual regret minimization (CFR) \cite{zinkevich2007regret}, whose details are introduced as follows, with the same notations as in extensive-form games in Section \ref{subs2.1}.

Recalling that $p^{\pi}(h)$ denotes the reach probability of history $h$ with strategy profile $\pi$. For an infoset $J\in I$, let $p^{\pi}(J)$ denote the probability of reaching the infoset $J$ via all possible histories in $J$, i.e., $p^{\pi}(J)=\sum_{h\in J}p^{\pi}(h)$. And denote by $p_i^{\pi}(J)$ the reach probability of infoset $J$ for player $i$ according to the strategy $\pi$, i.e., $p_i^{\pi}(J)=\Pi_{J'\cdot a'\sqsubseteq J,P(J)=i}p(J',a')$, and $p_{-i}^{\pi}(J)$ the counterfactual reach probability of infoset $J$, i.e., the probability of reaching $J$ with strategy profile $\pi$ except that the probability of reaching $J$ is treated as $1$ by the current actions of player $i$, i.e., without the contribution of player $i$ to reach $J$. Meanwhile, $p^{\pi}(h,z)$ denotes the probability of going from history $h$ to a nonterminal node $z\in Z$. Then, for player $i\in[n]$, the {\em counterfactual value} at a nonterminal history $h$ is defined as
\begin{align}
\nu_i^{\pi}(h):=\sum_{z\in Z,h\sqsubset z}p_{-i}^{\pi}(h)p^{\pi}(h,z)u_i(z),         \label{al-zs5}
\end{align}
the {\em counterfactual value} of an infoset $J$ is defined as
\begin{align}
\nu_i^{\pi}(J):=\sum_{h\in J}\nu_i^{\pi}(h),                                         \label{al-zs6}
\end{align}
and the {\em counterfactual value} of an action $a$ is defined as
\begin{align}
\nu_i^{\pi}(J,a):=\sum_{h\in J}\Big[p_{-i}^{\pi}(h)\sum_{z\in Z}p^{\pi}(h\cdot a,z)u_i(z)\Big].         \label{al-zs7}
\end{align}
The {\em instantaneous regret} at iteration $t$ and {\em counterfactual regret} at iteration $T$ for action $a$ in infoset $J$ are, respectively, defined as
\begin{align}
r_i^t(J,a)&:=\nu_i^{\pi^t}(J,a)-\nu_i^{\pi^t}(J),           \label{al-zs8}\\
R_i^T(J,a)&:=\sum_{t=1}^T r_i^t(J,a),                     \label{al-zs9}
\end{align}
where $\pi^t$ is the joint strategy profile leveraged at iteration $t$.

By defining $R_i^{T,+}(J,a):=\max\{R_i^T(J,a),0\}$, applying regret matching by Hart and Mas-Colell \cite{hart2000simple} can generate the strategy update as
\begin{align}
\pi_i^{T+1}(J,a)=\left\{
                   \begin{array}{ll}
                     \frac{R_i^{T,+}(J,a)}{\zeta_i^T(J,a)}, & \text{if~} \zeta_i^T(J,a)>0 \\
                     \frac{1}{|A(J)|}, & \text{otherwise}
                   \end{array}
                 \right.                                    \label{al-zs10}
\end{align}
with $\zeta_i^T(J,a):=\sum_{a\in A(J)}R_i^{T,+}(J,a)$, and (\ref{al-zs10}) is the essential CFR method for player $i$'s strategy selection. Moreover, it is known that the CFR method can guarantee the convergence to NEs for the average strategy of players, i.e.,
\begin{align}
\bar{\pi}_i^T(J,a):=\frac{\sum_{t=1}^T p_i^{\pi^t}(J)\pi_i^t(J,a)}{\sum_{t=1}^T p_i^{\pi^t}(J)},~~~\forall i\in[n].         \label{al-zs11}
\end{align}

Hitherto, various famous variants of CFR have been developed with superior performance, including CFR+ \cite{tammelin2014solving,bowling2015heads}, discounted CFR (DCFR) \cite{brown2019solving}, linear CFR (LCFR) \cite{brown2019deep}, exponential CFR (ECFR) \cite{li2020solving}, AutoCFR \cite{xu2022auto}, etc. More details can be found in \cite{neller2013introduction,li2021survey,brown2020equilibrium}.

Meanwhile, lots of AI methods have been brought forward in the literature \cite{gidel2020multi}, such as policy space response oracles (PSRO) \cite{lanctot2017unified,muller2020generalized}, neural fictitious self-play \cite{heinrich2016deep}, deep CFR \cite{brown2019deep}, single deep CFR \cite{steinberger2019single}, unified deep equilibrium finding (UDEF) \cite{wang2022unified}, player of games (PoG) \cite{schmid2021player}, neural auto-curricula (NAC) \cite{feng2021neural}, and so forth. Among these methods, PSRO has been an effective approach in recent years, which unifies fictitious play and double oracle algorithms. Nonetheless, UDEF provides a unified framework of PSRO and CFR, which are generally considered independently with their own advantages, and thus UDEF are superior to both PSRO and CFR as demonstrated by experiments on Leduc poker \cite{wang2022unified}. The recently-developed PoG algorithm has unified several previous approaches by integrating guided search, self-play learning, and game-theoretic reasoning, and demonstrated theoretically and experimentally the achievement of strong empirical performance in large perfect and imperfect information games, which defeats state-of-the-art in heads-up no-limit Texas Hold'em poker (Slumbot) \cite{schmid2021player}. Moreover, NAC, as a meta-learning algorithm proposed recently in \cite{feng2021neural}, provides a potential future direction to develop general multi-agent reinforcement learning (MARL) algorithms solely from data, since it can learn its own objective solely from the interactions with environment, without the need of human-designed knowledge about game theoretic principles, and it can decide by itself what the meta-solution, i.e., who to compete with, should be during training. Furthermore, it is shown that NAC is comparable or even superior to the state-of-the-art population-based game solvers, such as PSRO, on a series of games, like Games of Skill, differentiable Lotto, non-transitive Mixture Games, Iterated Matching Pennies, and Kuhn poker \cite{feng2021neural}.

Finally, it is worth pointing out that by CFR methods, it can guarantee the convergence to NEs in the sense of the empirical distribution (i.e., time-average) of play, but generally failing to converge for the day-to-day play (i.e., the last-iterate convergence) \cite{mertikopoulos2018cycles,vlatakis2020no}, although it does converge in the sense of last-iterate in two-player zero-sum games \cite{anagnostides2022last}. In this respect, the last-iterate convergence is of also importance to be explored as demonstrated in economics, and so on \cite{daskalakis2018last,abernethy2019fictitious,golowich2020last,wei2020linear,anagnostides2022last}.

\subsection{Stackelberg Games}\label{sec4.2}

GSGs and SSGs can be expressed as bilevel linear programming (BLP) or mixed integer linear programming (MILP), which can be further transformed or relaxed as linear programming (LP) \cite{casorran2017formulations}. As mentioned in Section \ref{sec3.2}, solving GSGs and SSGs is generally {\sf NP}-hard, and most existing solution methods are variants of solution approaches for MILP and LP, including cutting plane methods, enumerative methods, hybrid methods, and so on \cite{arriagada2021benders}. Some of most widely used approaches in the literature are introduced in the sequel.

\begin{enumerate}
  \item {\em Multiple LP Approach.} This approach is proposed in \cite{conitzer2006computing}, most widely employed for those easy problems that can be solved in polynomial time, including the case with a single follower type for GSGs \cite{conitzer2006computing}, further improved upon in \cite{conitzer2011commitment} by merging LPs into a single MILP. And this approach has also been improved to deal with SSGs in \cite{korzhyk2010complexity}, generally pretty efficient in the case with size $1$ of the schedule and the case with size $2$ of the schedule but for homogeneous resources, as shown in Table \ref{tbsg1}.
  \item {\em Benders Decomposition.} Benders decomposition method is developed in \cite{bnnobrs1962partitioning}, which is effective to handle general MILP problems. The crux of this method is to divide the original problem into two other problems, that is, one is called master problem by relaxing some constraints and the other is called subproblem, along with a separation problem that is the dual of the subproblem. Then, the solution seeking procedure involves the solving of the master problem firstly, followed by solving the separation problem, and finally checking the feasibility and optimality conditions for the subproblem with different contingent operations. Moreover, this approach can be improved upon by combining with other techniques, such as Farkas' lemma \cite{farkas1902theorie} and normalized cut \cite{fischetti2009minimal}, leading to a recent efficient algorithm, called normalized Benders decomposition \cite{arriagada2021benders}, etc.
  \item {\em Branch and Cut.} Branch \& cut method, as a hybrid methods, combines the cutting plane method \cite{gomory1958outline} with the branch and bound method \cite{land1960automatic}. This approach is pretty effective for resolving various (mixed) integer programming problems while still ensuring the optimality. In general, branch and cut algorithm is in the same spirit of the branch and bound scheme, but appending new constraints when necessary in each node by resorting to cutting plane approaches \cite{arriagada2021benders}.
  \item {\em Cut and Branch.} This method is similar to the branch and cut approach, and the difference lies in that the extra cuts are only added in the root node. Meanwhile, only the branching constraints are added to the other nodes. It is found in \cite{arriagada2021benders} that with variables in $\mathbb{R}$ in master problem and stabilization, cut and branch is superior to other methods in some sense.
  \item {\em Gradient Descent Ascent.} Gradient descent ascent, i.e., the classical gradient descent and ascent algorithm \cite{ruder2016overview}, is the most noticeable algorithm for solving continuous Stackelberg games, where descent and ascent operations are, respectively, performed for the leader and the follower, but in a sequential order, and other methods mostly rest on this algorithm \cite{goktas2021convex,fiez2019convergence}. For example, the max-oracle gradient-descent algorithm \cite{goktas2021convex} is a variant of gradient descent ascent, where the ascent operation in the follower is directly replaced with an approximate best response provided by a max-oracle.
\end{enumerate}

Finally, it is worth pointing out that AI methods have also been leveraged to cope with Stackelberg games, e.g., \cite{gottipati2021genetic} and a survey \cite{de2016machine} for reference.

\subsection{Zero-Sum Differential Games}\label{sec4.3}

Among the methods for solving zero-sum DGs, the viscosity solution approach is the most widely exploited one, for which it is known that a value function is the solution of the Hamilton-Jacobi-Isaacs (HJI) equations. In the sequel, this approach is introduced for DGs (\ref{10}) and (\ref{11}), and other detailed cases can be found in \cite{friedman2013differential,tran2021hamilton}.

For DGs (\ref{10}) and (\ref{11}), the Hamiltonian is defined as
\begin{align}
H(t,x,\omega)&=\min_{u\in U}\max_{v\in V}\{\langle f(t,x,u,v),\omega\rangle+f_0(t,x,u,v)\},      \nonumber\\
&\hspace{2.6cm} t\in[t_0,T],~~x,\omega\in\mathbb{R}^d                                    \label{adg1}
\end{align}
and the HJI equation is given as
\begin{align}
&\partial_t \psi(t,z)+H(t,z,\partial_z \psi(t,z))=0,         \nonumber\\
&\psi(T,z)=\phi(z),~~~t\in[t_0,T),~z\in\mathbb{R}^d                      \label{adg2}
\end{align}
where the second condition is called the terminal condition, $\psi:[t_0,T]\times\mathbb{R}^d\to\mathbb{R}$ is a function, and $\partial_t,\partial_z$ represent the subgradients with respect to $t,z$, respectively.

\begin{figure}[h]
\centering
\includegraphics[width=1.8in]{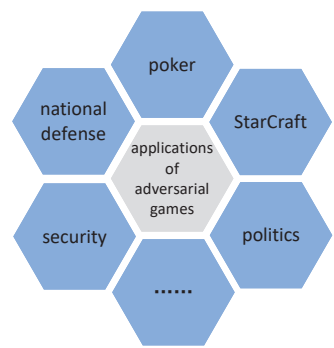}
\caption{A schematic illustration of applications of adversarial games.}
\label{application}
\end{figure}

Let $\Psi$ denote the set of functions $\psi:[t_0,T]\times\mathbb{R}^d\to\mathbb{R}$ satisfying the continuity condition in $t$ and the Lipschitz condition on every bounded subset of $\mathbb{R}^d$ in $x$. From \cite{plaksin2019hamilton}, it is known that if a function $\psi\in\Psi$ is coinvariantly differentiable at each point $(t,z)\in [t_0,T]\times \mathbb{R}^d$, satisfies HJI equation (\ref{adg2}), and $\partial_t\psi,\partial_z\psi\in\Psi$, then $\psi$ is the value function of differential game (\ref{10}) and (\ref{11}), and the optimal control strategies for two players are given as
\begin{align}
u^*(t,z)&\in\arg\min_{u\in U}\max_{v\in V} \chi(t,z,u,v),        \nonumber\\
v^*(t,z)&\in\arg\max_{v\in V}\min_{u\in U}\chi(t,z,u,v),        \label{adg3}
\end{align}
where
\begin{align}
\chi(t,z,u,v):=\langle f(t,z,u,v),\partial_z \psi(t,z)\rangle+f_0(t,z,u,v).     \label{adg4}
\end{align}

Moreover, it should be noted that AI methods have also been applied to solve differential games, for example, reinforcement learning was employed to deal with multi-player nonlinear differential games \cite{li2020multiplayer}, where a novel two-level value iteration-based integral reinforcement learning algorithm was proposed only depending upon partial information of system dynamics.

\section{Applications}\label{sec5}

This section provides some practical applications for adversarial games. As a matter of fact, adversarial games have been leveraged to solve a large volume of realistic problems in the literature, as illustrated in Fig. \ref{application}, including poker \cite{schmid2021player}, StarCraft \cite{ontanon2013survey}, politics \cite{davidai2019politics}, infrastructure security \cite{sinha2018stackelberg}, pursuit-evasion problems \cite{weintraub2020introduction}, border defense \cite{bucarey2017building,von2020multiple,shishika2020review}, national defense \cite{ho2022game}, communication scheduling \cite{gao2019communication}, autonomous driving \cite{na2020theoretical}, homeland security \cite{albert2022homeland}, etc. In what follows, we provide three well known examples to illustrate applications.

\begin{figure}[h]
\centering
\includegraphics[width=2.0in]{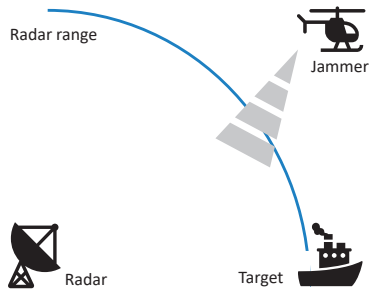}
\caption{A schematic illustration of radar jamming.}
\label{ap-zs1}
\end{figure}

\begin{example}[Radar Jamming]\label{ex1}
Radar jamming is one of widely studied applications of zero-sum games in modern electronic warfare \cite{song2011mimo,li2022counterfactual}. In radar jamming, there exist two players, one radar who aims to detect a target in a probability as high as possible, and one jammer who aims at minimizing the radar's  detection by jamming it. Therefore, the two players are diametrically opposed, and the scenario forms a two-player zero-sum game (cf. Fig. \ref{ap-zs1} for a schematic illustration). Usually, according to the type of the target, some kinds of utility functions can be constructed in distinct scenarios of jamming, and some constraints can be described mathematically relying on physical limitations, such as jammer power, spatial extent of jamming, and threshold parameter and reference window size for the radar. For example, a Swerling Type II target is assumed in \cite{bachmann2011game} in presence of Rayleigh distributed clutter, for which certain utility functions are built for cell averaging and order-statistic constant false alarm rate (CFAR) processors in three scenarios of jamming, i.e., ungated range noise, range-gated noise, and false-target jamming.
\end{example}

\begin{figure}[h]
\centering
\includegraphics[width=3.0in]{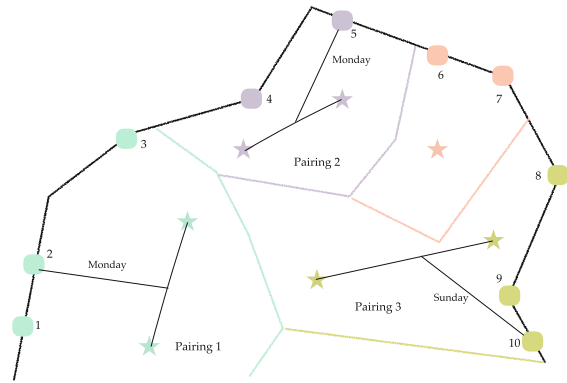}
\caption{Feasible schedule for a week, where stars and squares mean precinct headquarters and border outposts, respectively, cited from \cite{bucarey2017building}.}
\label{ap-ssg1}
\end{figure}

\begin{example}[Border Patrols]\label{ex2}
It is an important task for a country to secure national borders to avoid illicit behaviors of drugs, contraband, and stowaway, etc. In this spirit, border patrols are introduced here as one application of SSGs, which is proposed by Carabineros de Chile \cite{bucarey2017building,bucarey2021coordinating}, to thwart drug trafficking, contraband and illegal entry. To this end, both day and night shift patrols along the border are arranged by Carabineros according to distinct requirements.

The night shift patrols are specially focused on. To make it practically implementable, the region is partitioned into some police precincts, some of which are paired up when scheduling the patrol, because of the vast expanses and harsh landscape at the border and the manpower limitation. In addition, a set of vantage locations have been identified by Carabineros along the border of the region, which are suited for conducting surveillance with high-tech equipments, like heat sensors and night goggles. A night shift action means the deployment of a joint detail with personnel from two paired precincts to carry out vigilance overnight at the vantage locations within the realm of the paired precincts. Meanwhile, in consideration of logistical constraints, a joint detail is deployed for every precinct pair to a surveillance location once a week. Fig. \ref{ap-ssg1} illustrates the case with $3$ pairings, $7$ precincts and $10$ locations.
\end{example}

\begin{example}[Pursuit-Evasion Problems]\label{ex3}
Pursuit-evasion problems are one of prevalent applications of zero-sum DGs, which have been widely applied to many practical problems, such as surveillance and navigation, in robotics and aerospace and so forth. In pursuit-evasion problems, there usually exist a collection of pursuers and evaders (one pursuer and one evader in the simplest case) possibly with a moving target or stationary target set/area, and the pursuers aim to capture or intercept the evaders who have opposed objectives \cite{weintraub2020introduction}. As a concrete example, consider a case where there exists one pursuer (or defender) which protects a maritime coastline or border from the attacking by two slower aircraft (or evaders). The pursuer needs to sequentially pursue the evaders and strives to intercept them as far as possible from the coastline. Meanwhile, the two evaders can collaborate and strive to minimize their combined distance to the coastline before they are intercepted. For this problem, a regular solution was provided for the differential game in \cite{garcia2019strategies}.
\end{example}

\section{Possible Future Directions}\label{sec6}

In view of some challenges in adversary games, this section attempts to present potential research directions in future, as discussed in the sequel.

\begin{itemize}
  \item {\em Efficient Algorithms Design.} Even though a wide range of algorithms have been proposed in the literature, as introduced above, efficient, fast and optimal algorithms with limited computing, storage, and memory capabilities are still the overarching research directions in (adversarial) games and artificial intelligence, which are far from fully explored, including a plethora of scenarios, e.g., equilibrium computation \cite{zhang2022equilibrium}, real-time strategy (RTS) making \cite{lelis2021planning}, exploiting suboptimal opponents \cite{liu2022learning}, attack resiliency \cite{banik2021attack}, and so forth.
  \item {\em Last-Iterate Convergence.} In general, no-regret learning can guarantee the convergence of the empirical distribution of play (i.e., time-average convergence) for each player to the set of NEs. However, the last-iterate convergence fails in general \cite{mertikopoulos2018cycles,vlatakis2020no}, although restricted classes of games indeed have the last-iterate convergence by no-regret learning algorithms, such as two-player zero-sum games \cite{anagnostides2022last}. Note that the last-iterate convergence is important in many practical applications, for example, generative adversarial networks (GANs) \cite{wang2017generative} and economics \cite{daskalakis2021near}, which have been receiving a growing interest in recent years \cite{lee2021last}.
  \item {\em Imperfect Information.} Imperfect information, as a possible main feature of many practical adversarial games, inflicts a major challenge in adversarial games, which is still being actively under explored, although an array of works have focused on it, e.g., \cite{perolat2021poincare,schmid2021search}.
  \item {\em Large Games.} For adversarial games with large action spaces and/or infosets, practical limitations, such as limited computing resources, impose the need of efficient algorithms design amenable to implementation with limited computation, storage and even communication \cite{henderson2021cybered}.
  \item {\em Incomplete Information.} Incomplete information is another main hallmark of many adversarial games, which is one of challenge sources. Generally speaking, game uncertainties, such as parameter uncertainties, action outcome uncertainty, underlying world state uncertainty, can be subsumed in the category of incomplete information, and the main studied models are Bayesian and interval models \cite{costikyan2013uncertainty,insua2020advances,xu2021learning}.
  \item {\em Bounded Rationality.} Completely rational players are often assumed in the study of games. Nonetheless, irrational players naturally appear in practice, which has triggered an increasing interest in games with bounded rationality, e.g., behavior models such as lens-QR models, prospect theory inspired models and quantal response models \cite{kar2015game,caballero2021identifying,tsiotras2021bounded}.
  \item {\em Dynamic Environments.} Most of games have been investigated as static ones, i.e., with time-invariant game rules. However, due to possible dynamic characteristics of the environment within which players compete, online game (or time-varying game) is imperative for further attention in future, where each player's utility function is time-varying or even adversarial without any distribution assumptions \cite{lu2020online,meng2021decentralized,meng2022decentralized,zhang2022no,li2022survey}.
  \item {\em Hybrid Games.} It is known that many realistic adversarial games involve both continuous and discrete physical dynamics that govern players' motion or changing rules, which can be framed in the framework of hybrid games \cite{platzer2015differential,iyermodeling}. In this respect, how to combine the game theory with control dynamics is an important yet challenging research area.
  \item {\em AI in Games.} Recent years have witnessed great progress in the success of AI methods applied in games, which can integrate some advanced approaches of reinforcement learning, neural networks, meta-learning, and so on \cite{brown2020combining,li2020openholdem,fu2021actor,oh2021creating}. With the advent of modern high-tech and big-data complex missions, AI methods provide an effective manner to commit real-time strategies by solely exploiting offline or real-time streaming data \cite{yin2021ai}.
\end{itemize}

\section{Conclusion}\label{sec7}

Adversarial games play a significant role in practical applications, for which this survey provided a systematic overview on it from three main categories, i.e., zero-sum normal- and extensive-form games, Stackelberg (security) games, and zero-sum differential games. To this end, several distinct angles have been employed to anatomize adversarial games, ranging from game models, solution concepts, problem classification, research frontiers, prevailing algorithms and real-world applications to potential future directions. In general, this survey has attempted to summarize the past research in an intact manner, although the existing references are too vast to cover in its entirety. To our best knowledge, this survey is the first to present a systematic overview on adversarial games. Finally, future possible directions have been also discussed.


\begin{thebibliography}{100}

\bibitem{von1947theory}
J.~von Neumann and O.~Morgenstern, \emph{{Theory of Games and Economic
  Behavior}, 2nd ed}.\hskip 1em plus 0.5em minus 0.4em\relax Princeton
  University Press, 1947.

\bibitem{nash1950equilibrium}
J.~F. Nash, ``Equilibrium points in $n$-person games,'' \emph{Proceedings of
  the National Academy of Sciences}, vol.~36, no.~1, pp. 48--49, 1950.

\bibitem{nash1951non}
J.~Nash, ``Non-cooperative games,'' \emph{Annals of Mathematics}, vol.~54,
  no.~2, pp. 286--295, 1951.

\bibitem{fudenberg1991game}
D.~Fudenberg and J.~Tirole, \emph{Game Theory}.\hskip 1em plus 0.5em minus
  0.4em\relax MIT Press, 1991.

\bibitem{osborne1994course}
M.~J. Osborne and A.~Rubinstein, \emph{A Course in Game Theory}.\hskip 1em plus
  0.5em minus 0.4em\relax MIT Press, 1994.

\bibitem{bacsar2018handbook}
T.~Ba{\c{s}}ar and G.~Zaccour, \emph{Handbook of Dynamic Game Theory}.\hskip
  1em plus 0.5em minus 0.4em\relax Springer International Publishing, 2018.

\bibitem{aumann1995repeated}
R.~J. Aumann, M.~Maschler, and R.~E. Stearns, \emph{Repeated Games with
  Incomplete Information}.\hskip 1em plus 0.5em minus 0.4em\relax MIT Press,
  1995.

\bibitem{bard2013annual}
N.~Bard, J.~Hawkin, J.~Rubin, and M.~Zinkevich, ``The annual computer poker
  competition,'' \emph{AI Magazine}, vol.~34, no.~2, pp. 112--112, 2013.

\bibitem{nguyen2016towards}
T.~H. Nguyen, D.~Kar, M.~Brown, A.~Sinha, A.~X. Jiang, and M.~Tambe, ``Towards
  a science of security games,'' in \emph{Mathematical Sciences with
  Multidisciplinary Applications}, 2016, pp. 347--381.

\bibitem{silver2016mastering}
D.~Silver, A.~Huang, C.~J. Maddison, A.~Guez, L.~Sifre, G.~Van Den~Driessche,
  J.~Schrittwieser, I.~Antonoglou, V.~Panneershelvam, M.~Lanctot \emph{et~al.},
  ``{Mastering the game of Go with deep neural networks and tree search},''
  \emph{Nature}, vol. 529, no. 7587, pp. 484--489, 2016.

\bibitem{silver2017mastering}
D.~Silver, J.~Schrittwieser, K.~Simonyan, I.~Antonoglou, A.~Huang, A.~Guez,
  T.~Hubert, L.~Baker, M.~Lai, A.~Bolton \emph{et~al.}, ``{Mastering the game
  of Go without human knowledge},'' \emph{Nature}, vol. 550, no. 7676, pp.
  354--359, 2017.

\bibitem{silver2018general}
D.~Silver, T.~Hubert, J.~Schrittwieser, I.~Antonoglou, M.~Lai, A.~Guez,
  M.~Lanctot, L.~Sifre, D.~Kumaran, T.~Graepel \emph{et~al.}, ``{A general
  reinforcement learning algorithm that masters chess, shogi, and Go through
  self-play},'' \emph{Science}, vol. 362, no. 6419, pp. 1140--1144, 2018.

\bibitem{sinha2018stackelberg}
A.~Sinha, F.~Fang, B.~An, C.~Kiekintveld, and M.~Tambe, ``{Stackelberg security
  games: Looking beyond a decade of success},'' in \emph{International Joint
  Conference on Artificial Intelligence (IJCAI)}, Stockholm, Sweden, 2018, pp.
  5494--5501.

\bibitem{li2021survey}
H.~Li, X.~Wang, F.~Jia, Y.~Li, and Q.~Chen, ``{A survey of Nash equilibrium
  strategy solving based on CFR},'' \emph{Archives of Computational Methods in
  Engineering}, vol.~28, no.~4, pp. 2749--2760, 2021.

\bibitem{sohrabi2020survey}
M.~K. Sohrabi and H.~Azgomi, ``A survey on the combined use of optimization
  methods and game theory,'' \emph{Archives of Computational Methods in
  Engineering}, vol.~27, no.~1, pp. 59--80, 2020.

\bibitem{wang2022cooperative}
J.~Wang, Y.~Hong, J.~Wang, J.~Xu, Y.~Tang, Q.-L. Han, and J.~Kurths,
  ``{Cooperative and competitive multi-agent systems: From optimization to
  games},'' \emph{IEEE/CAA Journal of Automatica Sinica}, vol.~9, no.~5, pp.
  763--783, 2022.

\bibitem{li2022survey}
X.~Li, L.~Xie, and N.~Li, ``A survey of decentralized online learning,''
  \emph{arXiv preprint arXiv:2205.00473}, 2022.

\bibitem{ho2022game}
E.~Ho, A.~Rajagopalan, A.~Skvortsov, S.~Arulampalam, and M.~Piraveenan, ``{Game
  theory in defence applications: A review},'' \emph{Sensors}, vol.~22, no.~3,
  p. 1032, 2022.

\bibitem{shishika2020review}
D.~Shishika and V.~Kumar, ``A review of multi-agent perimeter defense games,''
  in \emph{International Conference on Decision and Game Theory for Security},
  College Park, USA, 2020, pp. 472--485.

\bibitem{zhu2021survey}
M.~Zhu, A.~H. Anwar, Z.~Wan, J.-H. Cho, C.~A. Kamhoua, and M.~P. Singh, ``{A
  survey of defensive deception: Approaches using game theory and machine
  learning},'' \emph{IEEE Communications Surveys \& Tutorials}, vol.~23, no.~4,
  pp. 2460--2493, 2021.

\bibitem{lanctot2017unified}
M.~Lanctot, V.~Zambaldi, A.~Gruslys, A.~Lazaridou, K.~Tuyls, J.~P{\'e}rolat,
  D.~Silver, and T.~Graepel, ``A unified game-theoretic approach to multiagent
  reinforcement learning,'' in \emph{Advances in Neural Information Processing
  Systems}, vol.~30, Long Beach, CA, USA, 2017.

\bibitem{littman1994markov}
M.~L. Littman, ``Markov games as a framework for multi-agent reinforcement
  learning,'' in \emph{Machine Learning Proceedings}, 1994, pp. 157--163.

\bibitem{zamir2008bayesian}
S.~Zamir \emph{et~al.}, ``{Bayesian games: Games with incomplete
  information},'' Tech. Rep., 2008.

\bibitem{chen2009settling}
X.~Chen, X.~Deng, and S.-H. Teng, ``{Settling the complexity of computing
  two-player Nash equilibria},'' \emph{Journal of the ACM (JACM)}, vol.~56,
  no.~3, pp. 1--57, 2009.

\bibitem{daskalakis2009complexity}
C.~Daskalakis, P.~W. Goldberg, and C.~H. Papadimitriou, ``{The complexity of
  computing a Nash equilibrium},'' \emph{SIAM Journal on Computing}, vol.~39,
  no.~1, pp. 195--259, 2009.

\bibitem{rubinstein2019hardness}
A.~Rubinstein, \emph{Hardness of Approximation Between P and NP}.\hskip 1em
  plus 0.5em minus 0.4em\relax Morgan \& Claypool, 2019.

\bibitem{aumann1974subjectivity}
R.~J. Aumann, ``Subjectivity and correlation in randomized strategies,''
  \emph{Journal of Mathematical Economics}, vol.~1, no.~1, pp. 67--96, 1974.

\bibitem{hannan1957approximation}
J.~Hannan, ``{Approximation to Bayes risk in repeated play},''
  \emph{Contributions to the Theory of Games}, vol.~3, no.~2, pp. 97--139,
  1957.

\bibitem{neumann1928theorie}
J.~V. Neumann, ``Zur theorie der gesellschaftsspiele,'' \emph{Mathematische
  Annalen}, vol. 100, no.~1, pp. 295--320, 1928.

\bibitem{farina2020coarse}
G.~Farina, T.~Bianchi, and T.~Sandholm, ``Coarse correlation in extensive-form
  games,'' in \emph{AAAI Conference on Artificial Intelligence}, vol.~34,
  no.~2, 2020, pp. 1934--1941.

\bibitem{celli2019computing}
A.~Celli, S.~Coniglio, and N.~Gatti, ``Computing optimal coarse correlated
  equilibria in sequential games,'' \emph{arXiv preprint arXiv:1901.06221},
  2019.

\bibitem{celli2018computational}
A.~Celli and N.~Gatti, ``Computational results for extensive-form adversarial
  team games,'' in \emph{AAAI Conference on Artificial Intelligence}, vol.~32,
  no.~1, 2018.

\bibitem{von1997team}
B.~von Stengel and D.~Koller, ``Team-maxmin equilibria,'' \emph{Games and
  Economic Behavior}, vol.~21, no. 1-2, pp. 309--321, 1997.

\bibitem{omidshafiei2019alpha}
S.~Omidshafiei, C.~Papadimitriou, G.~Piliouras, K.~Tuyls, M.~Rowland, J.-B.
  Lespiau, W.~M. Czarnecki, M.~Lanctot, J.~Perolat, and R.~Munos,
  ``{$\alpha$-rank: Multi-agent evaluation by evolution},'' \emph{Scientific
  Reports}, vol.~9, no.~1, pp. 1--29, 2019.

\bibitem{von1934marktform}
H.~Von~Stackelberg, \emph{Marktform und gleichgewicht}.\hskip 1em plus 0.5em
  minus 0.4em\relax Springer-Verlag, Berlin, 1934.

\bibitem{an2013deployed}
B.~An, F.~Ord{\'o}{\~n}ez, M.~Tambe, E.~Shieh, R.~Yang, C.~Baldwin,
  J.~DiRenzo~III, K.~Moretti, B.~Maule, and G.~Meyer, ``{A deployed quantal
  response-based patrol planning system for the U.S. coast guard},''
  \emph{Interfaces}, vol.~43, no.~5, pp. 400--420, 2013.

\bibitem{casorran2019study}
C.~Casorr{\'a}n, B.~Fortz, M.~Labb{\'e}, and F.~Ord{\'o}{\~n}ez, ``{A study of
  general and security Stackelberg game formulations},'' \emph{European Journal
  of Operational Research}, vol. 278, no.~3, pp. 855--868, 2019.

\bibitem{conitzer2006computing}
V.~Conitzer and T.~Sandholm, ``Computing the optimal strategy to commit to,''
  in \emph{Proceedings of the 7th ACM conference on Electronic Commerce},
  Michigan, USA, 2006, pp. 82--90.

\bibitem{leitmann1978generalized}
G.~Leitmann, ``{On generalized Stackelberg strategies},'' \emph{Journal of
  Optimization Theory and Applications}, vol.~26, no.~4, pp. 637--643, 1978.

\bibitem{von2010market}
H.~von Stackelberg, \emph{Market Structure and Equilibrium}.\hskip 1em plus
  0.5em minus 0.4em\relax Springer Science \& Business Media, 2011.

\bibitem{bacsar1998dynamic}
T.~Ba{\c{s}}ar and G.~J. Olsder, \emph{Dynamic Noncooperative Game
  Theory}.\hskip 1em plus 0.5em minus 0.4em\relax SIAM, 1998.

\bibitem{issacs1965diff}
R.~Isaacs, \emph{Differential Games}.\hskip 1em plus 0.5em minus 0.4em\relax
  Wiley, New York, 1965.

\bibitem{lewis2012optimal}
F.~L. Lewis, D.~Vrabie, and V.~L. Syrmos, \emph{Optimal Control}.\hskip 1em
  plus 0.5em minus 0.4em\relax John Wiley \& Sons, 2012.

\bibitem{buckdahn2011some}
R.~Buckdahn, P.~Cardaliaguet, and M.~Quincampoix, ``Some recent aspects of
  differential game theory,'' \emph{Dynamic Games and Applications}, vol.~1,
  no.~1, pp. 74--114, 2011.

\bibitem{friedman2013differential}
A.~Friedman, \emph{Differential Games}.\hskip 1em plus 0.5em minus 0.4em\relax
  Courier Corporation, 2013.

\bibitem{garg2011bilinear}
J.~Garg, A.~X. Jiang, and R.~Mehta, ``{Bilinear games: Polynomial time
  algorithms for rank based subclasses},'' in \emph{International Workshop on
  Internet and Network Economics}, Singapore, 2011, pp. 399--407.

\bibitem{lemke1964equilibrium}
C.~E. Lemke and J.~T. Howson, Jr, ``Equilibrium points of bimatrix games,''
  \emph{Journal of the Society for Industrial and Applied Mathematics},
  vol.~12, no.~2, pp. 413--423, 1964.

\bibitem{anagnostides2020solving}
I.~Anagnostides and P.~Penna, ``Solving zero-sum games through alternating
  projections,'' \emph{arXiv preprint arXiv:2010.00109}, 2021.

\bibitem{dinh2021online}
L.~C. Dinh, Y.~Yang, Z.~Tian, N.~P. Nieves, O.~Slumbers, D.~H. Mguni, H.~B.
  Ammar, and J.~Wang, ``Online double oracle,'' \emph{arXiv preprint
  arXiv:2103.07780}, 2021.

\bibitem{murhekar2020approximate}
A.~Murhekar, ``{Approximate Nash equilibria of imitation games: Algorithms and
  complexity},'' in \emph{International Conference on Autonomous Agents and
  Multiagent Systems}, 2020, pp. 887--894.

\bibitem{borel1921theorie}
E.~Borel, ``La th{\'e}orie du jeu et les {\'e}quations int{\'e}gralesa noyau
  sym{\'e}trique,'' \emph{Comptes rendus de l'Acad{\'e}mie des Sciences}, vol.
  173, no. 1304-1308, p.~58, 1921.

\bibitem{howson1972equilibria}
J.~T. Howson~Jr, ``Equilibria of polymatrix games,'' \emph{Management Science},
  vol.~18, no. 5-part-1, pp. 312--318, 1972.

\bibitem{sengodan2020linear}
G.~Sengodan and C.~Arumugasamy, ``Linear complementarity problems and bilinear
  games,'' \emph{Applications of Mathematics}, vol.~65, no.~5, pp. 665--675,
  2020.

\bibitem{deligkas2022polynomial}
A.~Deligkas, M.~Fasoulakis, and E.~Markakis, ``{A polynomial-time algorithm for
  $1/3$-approximate Nash equilibria in bimatrix games},'' \emph{arXiv preprint
  arXiv:2204.11525}, 2022.

\bibitem{deligkas2020tree}
A.~Deligkas, J.~Fearnley, and R.~Savani, ``{Tree polymatrix games are
  PPAD-hard},'' \emph{arXiv preprint arXiv:2002.12119}, 2020.

\bibitem{seddighin2019campaigning}
S.~Seddighin, ``{Campaigning via LPs: Solving Blotto and Beyond},'' Ph.D.
  dissertation, University of Maryland, College Park, 2019.

\bibitem{mehta2018constant}
R.~Mehta, ``{Constant rank two-player games are PPAD-hard},'' \emph{SIAM
  Journal on Computing}, vol.~47, no.~5, pp. 1858--1887, 2018.

\bibitem{boodaghians2020smoothed}
S.~Boodaghians, J.~Brakensiek, S.~B. Hopkins, and A.~Rubinstein, ``{Smoothed
  complexity of $2$-player Nash equilibria},'' in \emph{Annual Symposium on
  Foundations of Computer Science}, 2020, pp. 271--282.

\bibitem{behnezhad2019optimal}
S.~Behnezhad, A.~Blum, M.~Derakhshan, M.~Hajiaghayi, C.~H. Papadimitriou, and
  S.~Seddighin, ``{Optimal strategies of Blotto games: Beyond convexity},'' in
  \emph{Proceedings of ACM Conference on Economics and Computation}, Phoenix,
  AZ, USA, 2019, pp. 597--616.

\bibitem{behnezhad2022fast}
S.~Behnezhad, S.~Dehghani, M.~Derakhshan, M.~Hajiaghayi, and S.~Seddighin,
  ``{Fast and simple solutions of Blotto games},'' \emph{Operations Research},
  DOI: 10.1287/opre.2022.2261, 2022.

\bibitem{beaglehole2022efficient}
D.~Beaglehole, ``{An efficient approximation algorithm for the Colonel Blotto
  game},'' \emph{arXiv preprint arXiv:2201.10758}, 2022.

\bibitem{leon2021bandit}
V.~Leon and S.~R. Etesami, ``{Bandit learning for dynamic Colonel Blotto game
  with a budget constraint},'' \emph{arXiv preprint arXiv:2103.12833}, 2021.

\bibitem{vu2019approximate}
D.~Q. Vu, P.~Loiseau, and A.~Silva, ``{Approximate equilibria in generalized
  Colonel Blotto and generalized Lottery Blotto games},'' \emph{arXiv preprint
  arXiv:1910.06559}, 2019.

\bibitem{boix2021multiplayer}
E.~Boix-Adser{\`a}, B.~L. Edelman, and S.~Jayanti, ``{The multiplayer Colonel
  Blotto game},'' \emph{Games and Economic Behavior}, vol. 129, pp. 15--31,
  2021.

\bibitem{vlatakis2019poincare}
E.-V. Vlatakis-Gkaragkounis, L.~Flokas, and G.~Piliouras, ``Poincar{\'e}
  recurrence, cycles and spurious equilibria in gradient-descent-ascent for
  non-convex non-concave zero-sum games,'' in \emph{Advances in Neural
  Information Processing Systems}, vol.~32, Vancouver, BC, Canada, 2019, pp.
  1--12.

\bibitem{zhang2022near}
G.~Zhang, Y.~Wang, L.~Lessard, and R.~B. Grosse, ``Near-optimal local
  convergence of alternating gradient descent-ascent for minimax
  optimization,'' in \emph{International Conference on Artificial Intelligence
  and Statistics}, 2022, pp. 7659--7679.

\bibitem{hamedani2021primal}
E.~Y. Hamedani and N.~S. Aybat, ``A primal-dual algorithm with line search for
  general convex-concave saddle point problems,'' \emph{SIAM Journal on
  Optimization}, vol.~31, no.~2, pp. 1299--1329, 2021.

\bibitem{tominin2021accelerated}
V.~Tominin, Y.~Tominin, E.~Borodich, D.~Kovalev, A.~Gasnikov, and
  P.~Dvurechensky, ``On accelerated methods for saddle-point problems with
  composite structure,'' \emph{arXiv preprint arXiv:2103.09344}, 2021.

\bibitem{xie2021dippa}
G.~Xie, Y.~Han, and Z.~Zhang, ``{DIPPA: An improved method for bilinear saddle
  point problems},'' \emph{arXiv preprint arXiv:2103.08270}, 2021.

\bibitem{kovalev2021accelerated}
D.~Kovalev, A.~Gasnikov, and P.~Richt{\'a}rik, ``Accelerated primal-dual
  gradient method for smooth and convex-concave saddle-point problems with
  bilinear coupling,'' \emph{arXiv preprint arXiv:2112.15199}, 2021.

\bibitem{thekumparampil2022lifted}
K.~K. Thekumparampil, N.~He, and S.~Oh, ``Lifted primal-dual method for
  bilinearly coupled smooth minimax optimization,'' \emph{arXiv preprint
  arXiv:2201.07427}, 2022.

\bibitem{gidel2017frank}
G.~Gidel, T.~Jebara, and S.~Lacoste-Julien, ``{Frank-Wolfe algorithms for
  saddle point problems},'' in \emph{International Conference on Artificial
  Intelligence and Statistics}, Florida, USA, 2017, pp. 362--371.

\bibitem{chen2020efficient}
C.~Chen, L.~Luo, W.~Zhang, and Y.~Yu, ``Efficient projection-free algorithms
  for saddle point problems,'' in \emph{Advances in Neural Information
  Processing Systems}, vol.~33, 2020, pp. 10\,799--10\,808.

\bibitem{li2021complexity}
H.~Li, Y.~Tian, J.~Zhang, and A.~Jadbabaie, ``Complexity lower bounds for
  nonconvex-strongly-concave min-max optimization,'' in \emph{Advances in
  Neural Information Processing Systems}, vol.~34, 2021, pp. 1--13.

\bibitem{hsieh2021limits}
Y.-P. Hsieh, P.~Mertikopoulos, and V.~Cevher, ``{The limits of min-max
  optimization algorithms: Convergence to spurious non-critical sets},'' in
  \emph{International Conference on Machine Learning}, 2021, pp. 4337--4348.

\bibitem{wei2020linear}
C.-Y. Wei, C.-W. Lee, M.~Zhang, and H.~Luo, ``Linear last-iterate convergence
  in constrained saddle-point optimization,'' in \emph{International Conference
  on Learning Representations}, 2021, pp. 1--12.

\bibitem{bistritz2021no}
I.~Bistritz, Z.~Zhou, X.~Chen, N.~Bambos, and J.~Blanchet, ``No weighted-regret
  learning in adversarial bandits with delays,'' \emph{Journal of Machine
  Learning Research}, vol.~23, pp. 1--43, 2022.

\bibitem{fiez2021online}
T.~Fiez, R.~Sim, S.~Skoulakis, G.~Piliouras, and L.~Ratliff, ``Online learning
  in periodic zero-sum games,'' vol.~34, 2021, pp. 1--13.

\bibitem{gao2021convergence}
H.~Gao, X.~Wang, L.~Luo, and X.~Shi, ``On the convergence of stochastic
  compositional gradient descent ascent method,'' in \emph{International Joint
  Conference on Artificial Intelligence}, 2021, pp. 1--7.

\bibitem{beznosikov2021distributed}
A.~Beznosikov, G.~Scutari, A.~Rogozin, and A.~Gasnikov, ``Distributed
  saddle-point problems under data similarity,'' vol.~34, 2021.

\bibitem{vlatakis2021solving}
E.-V. Vlatakis-Gkaragkounis, L.~Flokas, and G.~Piliouras, ``Solving min-max
  optimization with hidden structure via gradient descent ascent,'' in
  \emph{Advances in Neural Information Processing Systems}, vol.~34, 2021, pp.
  1--14.

\bibitem{goktas2021convex}
D.~Goktas and A.~Greenwald, ``{Convex-concave min-max Stackelberg games},'' in
  \emph{Advances in Neural Information Processing Systems}, vol.~34, 2021.

\bibitem{xefteris2015symmetric}
D.~Xefteris, ``Symmetric zero-sum games with only asymmetric equilibria,''
  \emph{Games and Economic Behavior}, vol.~89, pp. 122--125, 2015.

\bibitem{cai2011minmax}
Y.~Cai and C.~Daskalakis, ``On minmax theorems for multiplayer games,'' in
  \emph{Proceedings of Annual ACM-SIAM Symposium on Discrete Algorithms}, San
  Francisco, California, 2011, pp. 217--234.

\bibitem{anagnostides2022last}
I.~Anagnostides, I.~Panageas, G.~Farina, and T.~Sandholm, ``On last-iterate
  convergence beyond zero-sum games,'' \emph{arXiv preprint arXiv:2203.12056},
  2022.

\bibitem{bailey2021left}
J.~P. Bailey, ``$o(1/t)$ time-average convergence in a generalization of
  multiagent zero-sum games,'' \emph{arXiv preprint arXiv:2110.02482}, 2021.

\bibitem{fiezonline}
T.~Fiez, R.~Sim, S.~Skoulakis, G.~Piliouras, and L.~Ratliff, ``{Online learning
  in periodic zero-sum games: von Neumann vs Poincar{\'e}}.''

\bibitem{skoulakis2021evolutionary}
S.~Skoulakis, T.~Fiez, R.~Sim, G.~Piliouras, and L.~Ratliff, ``{Evolutionary
  game theory squared: Evolving agents in endogenously evolving zero-sum
  games},'' in \emph{AAAI Conference on Artificial Intelligence}, 2021, pp.
  1--9.

\bibitem{hughes2020learning}
E.~Hughes, T.~W. Anthony, T.~Eccles, J.~Z. Leibo, D.~Balduzzi, and Y.~Bachrach,
  ``Learning to resolve alliance dilemmas in many-player zero-sum games,''
  \emph{arXiv preprint arXiv:2003.00799}, 2020.

\bibitem{ganzfried2020fast}
S.~Ganzfried, ``{Fast complete algorithm for multiplayer Nash equilibrium},''
  \emph{arXiv preprint arXiv:2002.04734}, 2020.

\bibitem{anagnostides2021near}
I.~Anagnostides, C.~Daskalakis, G.~Farina, M.~Fishelson, N.~Golowich, and
  T.~Sandholm, ``Near-optimal no-regret learning for correlated equilibria in
  multi-player general-sum games,'' \emph{arXiv preprint arXiv:2111.06008},
  2021.

\bibitem{anagnostides2022faster}
I.~Anagnostides, G.~Farina, C.~Kroer, A.~Celli, and T.~Sandholm, ``Faster
  no-regret learning dynamics for extensive-form correlated and coarse
  correlated equilibria,'' \emph{arXiv preprint arXiv:2202.05446}, 2022.

\bibitem{gidel2020multi}
G.~Gidel, ``Multi-player games in the era of machine learning,'' Ph.D.
  dissertation, Universit\'{e} de Montr\'{e}al, 2020.

\bibitem{zhang2020converging}
Y.~Zhang and B.~An, ``Converging to team-maxmin equilibria in zero-sum
  multiplayer games,'' in \emph{International Conference on Machine Learning},
  2020, pp. 11\,033--11\,043.

\bibitem{kalogiannis2021teamwork}
F.~Kalogiannis, E.-V. Vlatakis-Gkaragkounis, and I.~Panageas, ``{Teamwork makes
  von Neumann work: Min-max optimization in two-team zero-sum games},''
  \emph{arXiv preprint arXiv:2111.04178}, 2021.

\bibitem{hansen2008approximability}
K.~A. Hansen, T.~D. Hansen, P.~B. Miltersen, and T.~B. S{\o}rensen,
  ``Approximability and parameterized complexity of minmax values,'' in
  \emph{International Workshop on Internet and Network Economics}, 2008, pp.
  684--695.

\bibitem{borgs2010myth}
C.~Borgs, J.~Chayes, N.~Immorlica, A.~T. Kalai, V.~Mirrokni, and
  C.~Papadimitriou, ``The myth of the folk theorem,'' \emph{Games and Economic
  Behavior}, vol.~70, no.~1, pp. 34--43, 2010.

\bibitem{gharesifard2013distributed}
B.~Gharesifard and J.~Cort{\'e}s, ``{Distributed convergence to Nash equilibria
  in two-network zero-sum games},'' \emph{Automatica}, vol.~49, no.~6, pp.
  1683--1692, 2013.

\bibitem{lou2015nash}
Y.~Lou, Y.~Hong, L.~Xie, G.~Shi, and K.~H. Johansson, ``Nash equilibrium
  computation in subnetwork zero-sum games with switching communications,''
  \emph{IEEE Transactions on Automatic Control}, vol.~61, no.~10, pp.
  2920--2935, 2015.

\bibitem{huang2021no}
S.~Huang, J.~Lei, Y.~Hong, and U.~V. Shanbhag, ``No-regret distributed learning
  in two-network zero-sum games,'' in \emph{Proceedings of IEEE Conference on
  Decision and Control}, Austin, Texas, 2021, pp. 924--929.

\bibitem{zhang2020computing}
Y.~Zhang and B.~An, ``Computing team-maxmin equilibria in zero-sum multiplayer
  extensive-form games,'' in \emph{AAAI Conference on Artificial Intelligence},
  vol.~34, no.~02, 2020, pp. 2318--2325.

\bibitem{carminati2022public}
L.~Carminati, F.~Cacciamani, M.~Ciccone, and N.~Gatti, ``Public information
  representation for adversarial team games,'' \emph{arXiv preprint
  arXiv:2201.10377}, 2022.

\bibitem{farina2020faster}
G.~Farina, A.~Celli, N.~Gatti, and T.~Sandholm, ``Faster algorithms for optimal
  ex-ante coordinated collusive strategies in extensive-form zero-sum games,''
  \emph{arXiv preprint arXiv:2009.10061}, 2020.

\bibitem{zhang2021team}
B.~H. Zhang and T.~Sandholm, ``Team correlated equilibria in zero-sum
  extensive-form games via tree decompositions,'' \emph{arXiv preprint
  arXiv:2109.05284}, 2021.

\bibitem{tammelin2015solving}
O.~Tammelin, N.~Burch, M.~Johanson, and M.~Bowling, ``{Solving heads-up limit
  Texas Hold'em},'' in \emph{International Joint Conference on Artificial
  Intelligence}, 2015, pp. 645--652.

\bibitem{moravvcik2017deepstack}
M.~Morav{\v{c}}{\'\i}k, M.~Schmid, N.~Burch, V.~Lis{\`y}, D.~Morrill, N.~Bard,
  T.~Davis, K.~Waugh, M.~Johanson, and M.~Bowling, ``{DeepStack: Expert-level
  artificial intelligence in heads-up no-limit poker},'' \emph{Science}, vol.
  356, no. 6337, pp. 508--513, 2017.

\bibitem{brown2018superhuman}
N.~Brown and T.~Sandholm, ``{Superhuman AI for heads-up no-limit poker:
  Libratus beats top professionals},'' \emph{Science}, vol. 359, no. 6374, pp.
  418--424, 2018.

\bibitem{munos2020fast}
R.~Munos, J.~Perolat, J.-B. Lespiau, M.~Rowland, B.~De~Vylder, M.~Lanctot,
  F.~Timbers, D.~Hennes, S.~Omidshafiei, A.~Gruslys \emph{et~al.}, ``{Fast
  computation of Nash equilibria in imperfect information games},'' in
  \emph{International Conference on Machine Learning}, 2020, pp. 7119--7129.

\bibitem{farina2021better}
G.~Farina, C.~Kroer, and T.~Sandholm, ``{Better regularization for sequential
  decision spaces: Fast convergence rates for Nash, correlated, and team
  equilibria},'' \emph{arXiv preprint arXiv:2105.12954}, 2021.

\bibitem{brown2017safe}
N.~Brown and T.~Sandholm, ``Safe and nested subgame solving for
  imperfect-information games,'' in \emph{Advances in Neural Information
  Processing Systems}, vol.~30, 2017, pp. 1--11.

\bibitem{brown2018depth}
N.~Brown, T.~Sandholm, and B.~Amos, ``Depth-limited solving for
  imperfect-information games,'' in \emph{Advances in Neural Information
  Processing Systems}, vol.~31, 2018, pp. 1--12.

\bibitem{brown2020equilibrium}
N.~Brown, ``Equilibrium finding for large adversarial imperfect-information
  games,'' Ph.D. dissertation, Carnegie Mellon University, 2020.

\bibitem{marsland1986review}
T.~A. Marsland, ``A review of game-tree pruning,'' \emph{ICGA Journal}, vol.~9,
  no.~1, pp. 3--19, 1986.

\bibitem{sandholm2015solving}
T.~Sandholm, ``Solving imperfect-information games,'' \emph{Science}, vol. 347,
  no. 6218, pp. 122--123, 2015.

\bibitem{schmid2021search}
M.~Schmid, ``Search in imperfect information games,'' \emph{arXiv preprint
  arXiv:2111.05884}, 2021.

\bibitem{kovavrik2021fast}
V.~Kova{\v{r}}{\'\i}k, D.~Milec, M.~{\v{S}}ustr, D.~Seitz, and V.~Lis{\`y},
  ``{Fast algorithms for poker require modelling it as a sequential Bayesian
  game},'' \emph{arXiv preprint arXiv:2112.10890}, 2021.

\bibitem{farina2022kernelized}
G.~Farina, C.-W. Lee, H.~Luo, and C.~Kroer, ``{Kernelized multiplicative
  weights for $0/1$-polyhedral games: Bridging the gap between learning in
  extensive-form and normal-form games},'' \emph{arXiv preprint
  arXiv:2202.00237}, 2022.

\bibitem{meng2022generalized}
L.~Meng and Y.~Gao, ``Generalized bandit regret minimizer framework in
  imperfect information extensive-form game,'' \emph{arXiv preprint
  arXiv:2203.05920}, 2022.

\bibitem{bai2022near}
Y.~Bai, C.~Jin, S.~Mei, and T.~Yu, ``Near-optimal learning of extensive-form
  games with imperfect information,'' \emph{arXiv preprint arXiv:2202.01752},
  2022.

\bibitem{kozuno2021model}
T.~Kozuno, P.~M{\'e}nard, R.~Munos, and M.~Valko, ``{Model-free learning for
  two-player zero-sum partially observable Markov games with perfect recall},''
  \emph{arXiv preprint arXiv:2106.06279}, 2021.

\bibitem{brown2019superhuman}
N.~Brown and T.~Sandholm, ``{Superhuman AI for multiplayer poker},''
  \emph{Science}, vol. 365, no. 6456, pp. 885--890, 2019.

\bibitem{blair2019ai}
A.~Blair and A.~Saffidine, ``{AI surpasses humans at six-player poker},''
  \emph{Science}, vol. 365, no. 6456, pp. 864--865, 2019.

\bibitem{wu2019hierarchical}
B.~Wu, ``{Hierarchical macro strategy model for MOBA game AI},'' in \emph{AAAI
  Conference on Artificial Intelligence}, vol.~33, no.~1, 2019, pp. 1206--1213.

\bibitem{tian2020joint}
Y.~Tian, Q.~Gong, and Y.~Jiang, ``Joint policy search for multi-agent
  collaboration with imperfect information,'' in \emph{Advances in Neural
  Information Processing Systems}, vol.~33, 2020, pp. 19\,931--19\,942.

\bibitem{ganzfried2020parallel}
S.~Ganzfried, C.~Laughlin, and C.~Morefield, ``{Parallel algorithm for Nash
  equilibrium in multiplayer stochastic games with application to naval
  strategic planning},'' in \emph{International Conference on Distributed
  Artificial Intelligence}, 2020, pp. 1--13.

\bibitem{weilin2021imperfect}
Y.~Weilin, H.~Zhenzhen, L.~Junren, X.~Jiahui, J.~Xiang, C.~Shaofei, Z.~Wanpeng,
  and C.~Jing, ``{Imperfect information game in multiplayer no-limit Texas
  Hold'em based on mean approximation and deep CFVnet},'' in \emph{Proceedings
  of China Automation Congress}, 2021, pp. 2459--2466.

\bibitem{heinrich2016deep}
J.~Heinrich and D.~Silver, ``Deep reinforcement learning from self-play in
  imperfect-information games,'' \emph{arXiv preprint arXiv:1603.01121}, 2016.

\bibitem{li2019double}
H.~Li, K.~Hu, S.~Zhang, Y.~Qi, and L.~Song, ``Double neural counterfactual
  regret minimization,'' in \emph{International Conference on Learning
  Representations}, 2019, pp. 1--13.

\bibitem{farnia2020gans}
F.~Farnia and A.~Ozdaglar, ``{Do GANs always have Nash equilibria?}'' in
  \emph{International Conference on Machine Learning}, 2020, pp. 3029--3039.

\bibitem{gruslys2020advantage}
A.~Gruslys, M.~Lanctot, R.~Munos, F.~Timbers, M.~Schmid, J.~Perolat,
  D.~Morrill, V.~Zambaldi, J.-B. Lespiau, J.~Schultz \emph{et~al.}, ``The
  advantage regret-matching actor-critic,'' \emph{arXiv preprint
  arXiv:2008.12234}, 2020.

\bibitem{ye2020towards}
D.~Ye, G.~Chen, W.~Zhang, S.~Chen, B.~Yuan, B.~Liu, J.~Chen, Z.~Liu, F.~Qiu,
  H.~Yu \emph{et~al.}, ``{Towards playing full MOBA games with deep
  reinforcement learning},'' in \emph{Advances in Neural Information Processing
  Systems}, vol.~33, 2020, pp. 621--632.

\bibitem{ye2020mastering}
D.~Ye, Z.~Liu, M.~Sun, B.~Shi, P.~Zhao, H.~Wu, H.~Yu, S.~Yang, X.~Wu, Q.~Guo
  \emph{et~al.}, ``{Mastering complex control in MOBA games with deep
  reinforcement learning},'' in \emph{AAAI Conference on Artificial
  Intelligence}, vol.~34, no.~04, 2020, pp. 6672--6679.

\bibitem{schmid2021player}
M.~Schmid, M.~Moravcik, N.~Burch, R.~Kadlec, J.~Davidson, K.~Waugh, N.~Bard,
  F.~Timbers, M.~Lanctot, Z.~Holland \emph{et~al.}, ``Player of games,''
  \emph{arXiv preprint arXiv:2112.03178}, 2021.

\bibitem{phillips2021reinforcement}
P.~Phillips, ``Reinforcement learning in two-player zero-sum simultaneous
  action games,'' \emph{arXiv preprint arXiv:2110.04835}, 2021.

\bibitem{fu2021actor}
H.~Fu, W.~Liu, S.~Wu, Y.~Wang, T.~Yang, K.~Li, J.~Xing, B.~Li, B.~Ma, Q.~Fu
  \emph{et~al.}, ``Actor-critic policy optimization in a large-scale
  imperfect-information game,'' in \emph{International Conference on Learning
  Representations}, 2021, pp. 1--12.

\bibitem{wang2022unified}
X.~Wang, J.~Cerny, S.~Li, C.~Yang, Z.~Yin, H.~Chan, and B.~An, ``A unified
  perspective on deep equilibrium finding,'' \emph{arXiv preprint
  arXiv:2204.04930}, 2022.

\bibitem{feng2021neural}
X.~Feng, O.~Slumbers, Z.~Wan, B.~Liu, S.~McAleer, Y.~Wen, J.~Wang, and Y.~Yang,
  ``Neural auto-curricula in two-player zero-sum games,'' in \emph{Advances in
  Neural Information Processing Systems}, vol.~34, 2021.

\bibitem{feng2021discovering}
X.~Feng, O.~Slumbers, Y.~Yang, Z.~Wan, B.~Liu, S.~McAleer, Y.~Wen, and J.~Wang,
  ``Discovering multi-agent auto-curricula in two-player zero-sum games,''
  \emph{arXiv preprint arXiv:2106.02745}, 2021.

\bibitem{yin2021ai}
Q.~Yin, J.~Yang, W.~Ni, B.~Liang, and K.~Huang, ``{AI in games: Techniques,
  challenges and opportunities},'' \emph{arXiv preprint arXiv:2111.07631},
  2021.

\bibitem{celli2019learning}
A.~Celli, A.~Marchesi, T.~Bianchi, and N.~Gatti, ``Learning to correlate in
  multi-player general-sum sequential games,'' in \emph{Advances in Neural
  Information Processing Systems}, vol.~32, 2019.

\bibitem{celli2020no}
A.~Celli, A.~Marchesi, G.~Farina, and N.~Gatti, ``No-regret learning dynamics
  for extensive-form correlated equilibrium,'' in \emph{Advances in Neural
  Information Processing Systems}, vol.~33, 2020, pp. 7722--7732.

\bibitem{song2022sample}
Z.~Song, S.~Mei, and Y.~Bai, ``Sample-efficient learning of correlated
  equilibria in extensive-form games,'' \emph{arXiv preprint arXiv:2205.07223},
  2022.

\bibitem{wei2021last}
C.-Y. Wei, C.-W. Lee, M.~Zhang, and H.~Luo, ``{Last-iterate convergence of
  decentralized optimistic gradient descent/ascent in infinite-horizon
  competitive Markov games},'' in \emph{Annual Conference on Learning Theory},
  2021, pp. 4259--4299.

\bibitem{mao2022provably}
W.~Mao and T.~Ba{\c{s}}ar, ``{Provably efficient reinforcement learning in
  decentralized general-sum Markov games},'' \emph{Dynamic Games and
  Applications}, pp. 1--22, 2022.

\bibitem{insua2020advances}
D.~R. Insua, F.~Ruggeri, R.~Soyer, and S.~Wilson, ``{Advances in Bayesian
  decision making in reliability},'' \emph{European Journal of Operational
  Research}, vol. 282, no.~1, pp. 1--18, 2020.

\bibitem{casorran2017formulations}
C.~Casorr{\'a}n-Amilburu, ``{Formulations and algorithms for general and
  security Stackelberg games},'' Ph.D. dissertation, Universit{\'e} libre de
  Bruxelles; Universidad de Chile, 2017.

\bibitem{arriagada2021benders}
I.~A. Arriagada~Fritz, ``{Benders decomposition based algorithms for general
  and security Stackelberg games},'' Master's thesis, Universidad de Chile,
  2021.

\bibitem{dempe2018bilevel}
S.~Dempe, \emph{Bilevel Optimization: Theory, Algorithms and
  Applications}.\hskip 1em plus 0.5em minus 0.4em\relax TU Bergakademie
  Freiberg, 2018, vol.~3.

\bibitem{li2017review}
T.~Li and S.~P. Sethi, ``{A review of dynamic Stackelberg game models},''
  \emph{Discrete \& Continuous Dynamical Systems-B}, vol.~22, no.~1, p. 125,
  2017.

\bibitem{maharjan2013dependable}
S.~Maharjan, Q.~Zhu, Y.~Zhang, S.~Gjessing, and T.~Basar, ``{Dependable demand
  response management in the smart grid: A Stackelberg game approach},''
  \emph{IEEE Transactions on Smart Grid}, vol.~4, no.~1, pp. 120--132, 2013.

\bibitem{yu2015real}
M.~Yu and S.~H. Hong, ``{A real-time demand-response algorithm for smart grids:
  A Stackelberg game approach},'' \emph{IEEE Transactions on Smart Grid},
  vol.~7, no.~2, pp. 879--888, 2015.

\bibitem{yang2013coping}
D.~Yang, G.~Xue, J.~Zhang, A.~Richa, and X.~Fang, ``{Coping with a smart jammer
  in wireless networks: A Stackelberg game approach},'' \emph{IEEE Transactions
  on Wireless Communications}, vol.~12, no.~8, pp. 4038--4047, 2013.

\bibitem{guzman2021sequential}
C.~Guzman, J.~Riffo, C.~Telha, and M.~Van~Vyve, ``{A sequential Stackelberg
  game for dynamic inspection problems},'' \emph{European Journal of
  Operational Research}, 2021.

\bibitem{jiang2021iiot}
Y.~Jiang, Y.~Zhong, and X.~Ge, ``{IIoT data sharing based on blockchain: A
  multi-leader multi-follower Stackelberg game approach},'' \emph{IEEE Internet
  of Things Journal}, vol.~9, no.~6, pp. 4396--4410, 2021.

\bibitem{leyffer2010solving}
S.~Leyffer and T.~Munson, ``Solving multi-leader-common-follower games,''
  \emph{Optimisation Methods \& Software}, vol.~25, no.~4, pp. 601--623, 2010.

\bibitem{zhang2016multi}
H.~Zhang, Y.~Xiao, L.~X. Cai, D.~Niyato, L.~Song, and Z.~Han, ``{A multi-leader
  multi-follower Stackelberg game for resource management in LTE unlicensed},''
  \emph{IEEE Transactions on Wireless Communications}, vol.~16, no.~1, pp.
  348--361, 2016.

\bibitem{mallozzi2017multi}
L.~Mallozzi and R.~Messalli, ``Multi-leader multi-follower model with
  aggregative uncertainty,'' \emph{Games}, vol.~8, no.~3, p.~25, 2017.

\bibitem{tran2020resource}
T.~D. Tran and L.~B. Le, ``{Resource allocation for multi-tenant network
  slicing: A multi-leader multi-follower Stackelberg game approach},''
  \emph{IEEE Transactions on Vehicular Technology}, vol.~69, no.~8, pp.
  8886--8899, 2020.

\bibitem{castiglioni2021committing}
M.~Castiglioni, A.~Marchesi, and N.~Gatti, ``Committing to correlated
  strategies with multiple leaders,'' \emph{Artificial Intelligence}, vol. 300,
  p. 103549, 2021.

\bibitem{pita2010robust}
J.~Pita, M.~Jain, M.~Tambe, F.~Ord{\'o}nez, and S.~Kraus, ``{Robust solutions
  to Stackelberg games: Addressing bounded rationality and limited observations
  in human cognition},'' \emph{Artificial Intelligence}, vol. 174, no.~15, pp.
  1142--1171, 2010.

\bibitem{bai2021sample}
Y.~Bai, C.~Jin, H.~Wang, and C.~Xiong, ``{Sample-efficient learning of
  Stackelberg equilibria in general-sum games},'' in \emph{Advances in Neural
  Information Processing Systems}, vol.~34, 2021.

\bibitem{kiekintveld2009computing}
C.~Kiekintveld, M.~Jain, J.~Tsai, J.~Pita, F.~Ord{\'o}nez, and M.~Tambe,
  ``Computing optimal randomized resource allocations for massive security
  games,'' in \emph{Proceedings of International Conference on Autonomous
  Agents and Multiagent Systems}, Budapest, Hungary, 2009, pp. 689--696.

\bibitem{jain2010software}
M.~Jain, J.~Tsai, J.~Pita, C.~Kiekintveld, S.~Rathi, M.~Tambe, and
  F.~Ord{\'o}nez, ``{Software assistants for randomized patrol planning for the
  LAX airport police and the federal air marshal service},'' \emph{Interfaces},
  vol.~40, no.~4, pp. 267--290, 2010.

\bibitem{korzhyk2010complexity}
D.~Korzhyk, V.~Conitzer, and R.~Parr, ``{Complexity of computing optimal
  Stackelberg strategies in security resource allocation games},'' in
  \emph{AAAI conference on Artificial Intelligence}, Georgia, USA, 2010, pp.
  805--810.

\bibitem{fang2016green}
F.~Fang and T.~H. Nguyen, ``{Green security games: Apply game theory to
  addressing green security challenges},'' \emph{ACM SIGecom Exchanges},
  vol.~15, no.~1, pp. 78--83, 2016.

\bibitem{brown2016one}
M.~Brown, A.~Sinha, A.~Schlenker, and M.~Tambe, ``{One size does not fit all: A
  game-theoretic approach for dynamically and effectively screening for
  threats},'' in \emph{AAAI Conference on Artificial Intelligence}, vol.~30,
  no.~1, Arizona, USA, 2016.

\bibitem{zhang2016keeping}
C.~Zhang, S.~Gholami, D.~Kar, A.~Sinha, M.~Jain, R.~Goyal, and M.~Tambe,
  ``{Keeping pace with criminals: An extended study of designing patrol
  allocation against adaptive opportunistic criminals},'' \emph{Games}, vol.~7,
  no.~3, p.~15, 2016.

\bibitem{dasgupta2021adversary}
P.~Dasgupta, J.~B. Collins, and R.~Mittu, \emph{Adversary-Aware Learning
  Techniques and Trends in Cybersecurity}.\hskip 1em plus 0.5em minus
  0.4em\relax Springer, 2021.

\bibitem{galinkin2021information}
E.~Galinkin, ``Information security games: A survey,'' \emph{arXiv preprint
  arXiv:2103.12520}, 2021.

\bibitem{bucarey2017building}
V.~Bucarey, C.~Casorr{\'a}n, {\'O}.~Figueroa, K.~Rosas, H.~Navarrete, and
  F.~Ord{\'o}{\~n}ez, ``{Building real Stackelberg security games for border
  patrols},'' in \emph{International Conference on Decision and Game Theory for
  Security}, Vienna, Austria, 2017, pp. 193--212.

\bibitem{bucarey2021coordinating}
V.~Bucarey, C.~Casorr{\'a}n, M.~Labb{\'e}, F.~Ordo{\~n}ez, and O.~Figueroa,
  ``{Coordinating resources in Stackelberg security games},'' \emph{European
  Journal of Operational Research}, vol. 291, no.~3, pp. 846--861, 2021.

\bibitem{lou2015equilibrium}
J.~Lou and Y.~Vorobeychik, ``Equilibrium analysis of multi-defender security
  games,'' in \emph{International Joint Conference on Artificial Intelligence
  (IJCAI)}, Buenos Aires, Argentina, 2015, pp. 596--602.

\bibitem{mutzari2022robust}
D.~Mutzari, Y.~Aumann, and S.~Kraus, ``{Robust solutions for multi-defender
  Stackelberg security games},'' \emph{arXiv preprint arXiv:2204.14000}, 2022.

\bibitem{li2016catcher}
Y.~Li, V.~Conitzer, and D.~Korzhyk, ``Catcher-evader games,'' in
  \emph{International Joint Conference on Artificial Intelligence (IJCAI)}, New
  York, USA, 2016, pp. 329--337.

\bibitem{wang2019repeated}
B.~Wang, Y.~Zhang, Z.-H. Zhou, and S.~Zhong, ``{On repeated Stackelberg
  security game with the cooperative human behavior model for wildlife
  protection},'' \emph{Applied Intelligence}, vol.~49, no.~3, pp. 1002--1015,
  2019.

\bibitem{ma2021decision}
W.~Ma, W.~Liu, K.~McAreavey, X.~Luo, Y.~Jiang, J.~Zhan, and Z.~Chen, ``A
  decision support framework for security resource allocation under
  ambiguity,'' \emph{International Journal of Intelligent Systems}, vol.~36,
  no.~1, pp. 5--52, 2021.

\bibitem{fiez2019convergence}
T.~Fiez, B.~Chasnov, and L.~J. Ratliff, ``{Convergence of learning dynamics in
  Stackelberg games},'' \emph{arXiv preprint arXiv:1906.01217}, 2019.

\bibitem{kulkarni2015existence}
A.~A. Kulkarni and U.~V. Shanbhag, ``{An existence result for hierarchical
  Stackelberg v/s Stackelberg games},'' \emph{IEEE Transactions on Automatic
  Control}, vol.~60, no.~12, pp. 3379--3384, 2015.

\bibitem{goktas2022robust}
D.~Goktas, J.~Zhao, and A.~Greenwald, ``{Robust no-regret learning in min-max
  Stackelberg games},'' \emph{arXiv preprint arXiv:2203.14126}, 2022.

\bibitem{maffioli2019dealing}
M.~Maffioli, ``Dealing with partial information in follower's behavior
  identification,'' Master's thesis, Politecnico di Milano, 2019.

\bibitem{cheng2022single}
Z.~Cheng, G.~Chen, and Y.~Hong, ``{Single-leader-multiple-followers Stackelberg
  security game with hypergame framework},'' \emph{IEEE Transactions on
  Information Forensics and Security}, vol.~17, pp. 954--969, 2022.

\bibitem{birmpas2021optimally}
G.~Birmpas, J.~Gan, A.~Hollender, F.~J. Marmolejo-Coss{\'\i}o, N.~Rajgopal, and
  A.~A. Voudouris, ``{Optimally deceiving a learning leader in Stackelberg
  games},'' \emph{Journal of Artificial Intelligence Research}, vol.~72, pp.
  507--531, 2021.

\bibitem{lukes1971global}
D.~Lukes and D.~Russell, ``A global theory for linear-quadratic differential
  games,'' \emph{Journal of Mathematical Analysis and Applications}, vol.~33,
  no.~1, pp. 96--123, 1971.

\bibitem{engwerda2009linear}
J.~Engwerda, ``{Linear quadratic differential games: An overview},''
  \emph{Advances in Dynamic Games and Their Applications}, pp. 1--34, 2009.

\bibitem{shinar2008solvability}
J.~Shinar, V.~Turetsky, V.~Y. Glizer, and E.~Ianovsky, ``Solvability of
  linear-quadratic differential games associated with pursuit-evasion
  problems,'' \emph{International Game Theory Review}, vol.~10, no.~04, pp.
  481--515, 2008.

\bibitem{weintraub2020introduction}
I.~E. Weintraub, M.~Pachter, and E.~Garcia, ``An introduction to
  pursuit-evasion differential games,'' in \emph{Proceedings of American
  Control Conference (ACC)}, Denver, CO, USA, 2020, pp. 1049--1066.

\bibitem{gibali2021analytic}
A.~Gibali and O.~Kelis, ``An analytic and numerical investigation of a
  differential game,'' \emph{Axioms}, vol.~10, no.~2, p.~66, 2021.

\bibitem{huang2021defending}
Y.~Huang, J.~Chen, and Q.~Zhu, ``Defending an asset with partial information
  and selected observations: A differential game framework,'' in
  \emph{Proceedings of IEEE Conference on Decision and Control (CDC)}, Austin,
  Texas, 2021, pp. 2366--2373.

\bibitem{huang2021pursuit}
Y.~Huang and Q.~Zhu, ``A pursuit-evasion differential game with strategic
  information acquisition,'' \emph{arXiv preprint arXiv:2102.05469}, 2021.

\bibitem{garcia2020two}
E.~Garcia, D.~W. Casbeer, M.~Pachter, J.~W. Curtis, and E.~Doucette, ``A
  two-team linear quadratic differential game of defending a target,'' in
  \emph{Proceedings of American Control Conference (ACC)}, Denver, CO, USA,
  2020, pp. 1665--1670.

\bibitem{li2020mean}
X.~Li, J.~Shi, and J.~Yong, ``Mean-field linear-quadratic stochastic
  differential games in an infinite horizon,'' \emph{arXiv preprint
  arXiv:2007.06130}, 2020.

\bibitem{zhang2011iterative}
H.~Zhang, Q.~Wei, and D.~Liu, ``An iterative adaptive dynamic programming
  method for solving a class of nonlinear zero-sum differential games,''
  \emph{Automatica}, vol.~47, no.~1, pp. 207--214, 2011.

\bibitem{song2019robust}
R.~Song, J.~Li, and F.~L. Lewis, ``{Robust optimal control for disturbed
  nonlinear zero-sum differential games based on single NN and least
  squares},'' \emph{IEEE Transactions on Systems, Man, and Cybernetics:
  Systems}, vol.~50, no.~11, pp. 4009--4019, 2019.

\bibitem{lukoyanov2003functional}
N.~Y. Lukoyanov, ``{Functional Hamilton-Jacobi type equations with
  ci-derivatives in control problems with hereditary information},''
  \emph{Nonlinear Functional Analysis and Applications}, vol.~8, no.~4, pp.
  535--555, 2003.

\bibitem{plaksin2019hamilton}
A.~Plaksin, ``{On Hamilton-Jacobi-Bellman-Isaacs equation for time-delay
  systems},'' \emph{IFAC-PapersOnLine}, vol.~52, no.~18, pp. 138--143, 2019.

\bibitem{meng2020alinear}
W.~Meng and J.~Shi, ``{A linear quadratic stochastic Stackelberg differential
  game with time delay},'' \emph{arXiv preprint arXiv:2012.14145}, 2020.

\bibitem{gomoyunov2020dynamic}
M.~I. Gomoyunov, ``{Dynamic programming principle and Hamilton-Jacobi-Bellman
  equations for fractional-order systems},'' \emph{SIAM Journal on Control and
  Optimization}, vol.~58, no.~6, pp. 3185--3211, 2020.

\bibitem{moon2019zero}
J.~Moon and T.~Basar, ``{Zero-sum differential games on the Wasserstein
  space},'' \emph{arXiv preprint arXiv:1912.06084}, 2019.

\bibitem{liu2014multiperson}
D.~Liu and Q.~Wei, ``Multiperson zero-sum differential games for a class of
  uncertain nonlinear systems,'' \emph{International Journal of Adaptive
  Control and Signal Processing}, vol.~28, no. 3-5, pp. 205--231, 2014.

\bibitem{shi2020linear}
J.~Shi and G.~Wang, ``{A linear-quadratic Stackelberg differential game with
  mixed deterministic and stochastic controls},'' \emph{arXiv preprint
  arXiv:2004.00653}, 2020.

\bibitem{megahed2019stackelberg}
A.~E.-M.~A. Megahed, ``{The Stackelberg differential game for
  counter-terrorism},'' \emph{Quality \& Quantity}, vol.~53, no.~1, pp.
  207--220, 2019.

\bibitem{lee2021hamilton}
D.~Lee and C.~J. Tomlin, ``{Hamilton-Jacobi equations for two classes of
  state-constrained zero-sum games},'' \emph{arXiv preprint arXiv:2106.15006},
  2021.

\bibitem{elliott1981optimal}
R.~Elliott and M.~Davis, ``Optimal play in a stochastic differential game,''
  \emph{SIAM Journal on Control and Optimization}, vol.~19, no.~4, pp.
  543--554, 1981.

\bibitem{moon2018risk}
J.~Moon, T.~E. Duncan, and T.~Ba{\c{s}}ar, ``Risk-sensitive zero-sum
  differential games,'' \emph{IEEE Transactions on Automatic Control}, vol.~64,
  no.~4, pp. 1503--1518, 2018.

\bibitem{sun2021two}
J.~Sun, ``Two-person zero-sum stochastic linear-quadratic differential games,''
  \emph{SIAM Journal on Control and Optimization}, vol.~59, no.~3, pp.
  1804--1829, 2021.

\bibitem{li2021value}
J.~Li, W.~Li, and H.~Zhao, ``On the value of a general stochastic differential
  game with ergodic payoff,'' \emph{arXiv preprint arXiv:2106.15894}, 2021.

\bibitem{shi2017linear}
J.~Shi, G.~Wang, and J.~Xiong, ``{Linear-quadratic stochastic Stackelberg
  differential game with asymmetric information},'' \emph{Science China
  Information Sciences}, vol.~60, no.~9, pp. 1--15, 2017.

\bibitem{moon2021linear}
J.~Moon, ``{Linear-quadratic stochastic Stackelberg differential games for
  jump-diffusion systems},'' \emph{SIAM Journal on Control and Optimization},
  vol.~59, no.~2, pp. 954--976, 2021.

\bibitem{sun2021zero}
J.~Sun, H.~Wang, and J.~Wen, ``{Zero-sum Stackelberg stochastic
  linear-quadratic differential games},'' \emph{arXiv preprint
  arXiv:2109.14893}, 2021.

\bibitem{huang2021robust}
J.~Huang, S.~Wang, and Z.~Wu, ``{Robust Stackelberg differential game with
  model uncertainty},'' \emph{IEEE Transactions on Automatic Control}, DOI:
  10.1109/TAC.2021.3097549, 2021.

\bibitem{zheng2021stackelberg}
Y.~Zheng and J.~Shi, ``Stackelberg stochastic differential game with asymmetric
  noisy observations,'' \emph{International Journal of Control}, pp. 1--21,
  2021.

\bibitem{evans1984differential}
L.~C. Evans and P.~E. Souganidis, ``{Differential games and representation
  formulas for solutions of Hamilton-Jacobi-Isaacs equations},'' \emph{Indiana
  University Mathematics Journal}, vol.~33, no.~5, pp. 773--797, 1984.

\bibitem{altarovici2013general}
A.~Altarovici, O.~Bokanowski, and H.~Zidani, ``{A general Hamilton-Jacobi
  framework for non-linear state-constrained control problems},'' \emph{ESAIM:
  Control, Optimisation and Calculus of Variations}, vol.~19, no.~2, pp.
  337--357, 2013.

\bibitem{mitchell2005time}
I.~M. Mitchell, A.~M. Bayen, and C.~J. Tomlin, ``{A time-dependent
  Hamilton-Jacobi formulation of reachable sets for continuous dynamic
  games},'' \emph{IEEE Transactions on Automatic Control}, vol.~50, no.~7, pp.
  947--957, 2005.

\bibitem{margellos2011hamilton}
K.~Margellos and J.~Lygeros, ``{Hamilton-Jacobi formulation for reach-avoid
  differential games},'' \emph{IEEE Transactions on Automatic Control},
  vol.~56, no.~8, pp. 1849--1861, 2011.

\bibitem{fisac2015reach}
J.~F. Fisac, M.~Chen, C.~J. Tomlin, and S.~S. Sastry, ``Reach-avoid problems
  with time-varying dynamics, targets and constraints,'' in \emph{Proceedings
  of International Conference on Hybrid Systems: Computation and Control},
  Seattle, Washington, 2015, pp. 11--20.

\bibitem{asri2021deterministic}
B.~E. Asri and H.~Lalioui, ``Deterministic differential games in infinite
  horizon involving continuous and impulse controls,'' \emph{arXiv preprint
  arXiv:2107.03524}, 2021.

\bibitem{moon2020linear}
J.~Moon, ``Linear-quadratic mean-field stochastic zero-sum differential
  games,'' \emph{Automatica}, vol. 120, p. 109067, 2020.

\bibitem{sun2021mean}
J.~Sun, H.~Wang, and Z.~Wu, ``Mean-field linear-quadratic stochastic
  differential games,'' \emph{Journal of Differential Equations}, vol. 296, pp.
  299--334, 2021.

\bibitem{hart2000simple}
S.~Hart and A.~Mas-Colell, ``A simple adaptive procedure leading to correlated
  equilibrium,'' \emph{Econometrica}, vol.~68, no.~5, pp. 1127--1150, 2000.

\bibitem{tammelin2014solving}
O.~Tammelin, ``{Solving large imperfect information games using CFR+},''
  \emph{arXiv preprint arXiv:1407.5042}, 2014.

\bibitem{brown1951iterative}
G.~W. Brown, ``Iterative solution of games by fictitious play,'' \emph{Activity
  Analysis of Production and Allocation}, vol.~13, no.~1, pp. 374--376, 1951.

\bibitem{ganzfried2020fictitious}
S.~Ganzfried, ``Fictitious play outperforms counterfactual regret
  minimization,'' \emph{arXiv preprint arXiv:2001.11165}, 2020.

\bibitem{mcmahan2003planning}
H.~B. McMahan, G.~J. Gordon, and A.~Blum, ``Planning in the presence of cost
  functions controlled by an adversary,'' in \emph{International Conference on
  Machine Learning}, Washington, USA, 2003, pp. 536--543.

\bibitem{xu2020distributed}
X.~Xu and Q.~Zhao, ``{Distributed no-regret learning in multiagent systems:
  Challenges and recent developments},'' \emph{IEEE Signal Processing
  Magazine}, vol.~37, no.~3, pp. 84--91, 2020.

\bibitem{zhang2022equilibrium}
H.~Zhang, A.~Lerer, and N.~Brown, ``Equilibrium finding in normal-form games
  via greedy regret minimization,'' \emph{arXiv preprint arXiv:2204.04826},
  2022.

\bibitem{lu2020online}
K.~Lu, G.~Li, and L.~Wang, ``{Online distributed algorithms for seeking
  generalized Nash equilibria in dynamic environments},'' \emph{IEEE
  Transactions on Automatic Control}, vol.~66, no.~5, pp. 2289--2296, 2020.

\bibitem{meng2021decentralized}
M.~Meng, X.~Li, Y.~Hong, J.~Chen, and L.~Wang, ``Decentralized online learning
  for noncooperative games in dynamic environments,'' \emph{arXiv preprint
  arXiv:2105.06200}, 2021.

\bibitem{meng2022decentralized}
M.~Meng, X.~Li, and J.~Chen, ``{Decentralized Nash equilibria learning for
  online game with bandit feedback},'' \emph{arXiv preprint arXiv:2204.09467},
  2022.

\bibitem{zhang2022no}
M.~Zhang, P.~Zhao, H.~Luo, and Z.-H. Zhou, ``No-regret learning in time-varying
  zero-sum games,'' \emph{arXiv preprint arXiv:2201.12736}, 2022.

\bibitem{daskalakis2021near}
C.~Daskalakis, M.~Fishelson, and N.~Golowich, ``Near-optimal no-regret learning
  in general games,'' \emph{Advances in Neural Information Processing Systems},
  vol.~34, pp. 1--13, 2021.

\bibitem{hsieh2021adaptive}
Y.-G. Hsieh, K.~Antonakopoulos, and P.~Mertikopoulos, ``{Adaptive learning in
  continuous games: Optimal regret bounds and convergence to Nash
  equilibrium},'' in \emph{Annual Conference on Learning Theory}, 2021, pp.
  2388--2422.

\bibitem{zinkevich2007regret}
M.~Zinkevich, M.~Johanson, M.~Bowling, and C.~Piccione, ``Regret minimization
  in games with incomplete information,'' in \emph{Advances in Neural
  Information Processing Systems}, vol.~20, 2007, pp. 1--8.

\bibitem{bowling2015heads}
M.~Bowling, N.~Burch, M.~Johanson, and O.~Tammelin, ``Heads-up limit hold'em
  poker is solved,'' \emph{Science}, vol. 347, no. 6218, pp. 145--149, 2015.

\bibitem{brown2019solving}
N.~Brown and T.~Sandholm, ``Solving imperfect-information games via discounted
  regret minimization,'' in \emph{AAAI Conference on Artificial Intelligence},
  vol.~33, no.~01, 2019, pp. 1829--1836.

\bibitem{brown2019deep}
N.~Brown, A.~Lerer, S.~Gross, and T.~Sandholm, ``Deep counterfactual regret
  minimization,'' in \emph{International Conference on Machine Learning}, 2019,
  pp. 793--802.

\bibitem{li2020solving}
H.~Li, X.~Wang, S.~Qi, J.~Zhang, Y.~Liu, Y.~Wu, and F.~Jia, ``Solving
  imperfect-information games via exponential counterfactual regret
  minimization,'' \emph{arXiv preprint arXiv:2008.02679v2}, 2020.

\bibitem{xu2022auto}
H.~Xu, K.~Li, H.~Fu, Q.~Fu, and J.~Xing, ``{AutoCFR: Learning to design
  counterfactual regret minimization algorithms},'' in \emph{AAAI Conference on
  Artificial Intelligence}, 2022, pp. 1--8.

\bibitem{neller2013introduction}
T.~W. Neller and M.~Lanctot, ``An introduction to counterfactual regret
  minimization,'' in \emph{Proceedings of Model AI Assignments, The Fourth
  Symposium on Educational Advances in Artificial Intelligence}, vol.~11, 2013.

\bibitem{muller2020generalized}
P.~Muller, S.~Omidshafiei, M.~Rowland, K.~Tuyls, J.~Perolat, S.~Liu, D.~Hennes,
  L.~Marris, M.~Lanctot, E.~Hughes, Z.~Wang, G.~Lever, N.~Heess, T.~Graepel,
  and R.~Munos, ``A generalized training approach for multiagent learning,'' in
  \emph{International Conference on Learning Representations}, 2020, pp. 1--13.

\bibitem{steinberger2019single}
E.~Steinberger, ``Single deep counterfactual regret minimization,'' \emph{arXiv
  preprint arXiv:1901.07621}, 2019.

\bibitem{mertikopoulos2018cycles}
P.~Mertikopoulos, C.~Papadimitriou, and G.~Piliouras, ``Cycles in adversarial
  regularized learning,'' in \emph{Proceedings of Annual ACM-SIAM Symposium on
  Discrete Algorithms}, New Orleans, LA, USA, 2018, pp. 2703--2717.

\bibitem{vlatakis2020no}
E.-V. Vlatakis-Gkaragkounis, L.~Flokas, T.~Lianeas, P.~Mertikopoulos, and
  G.~Piliouras, ``{No-regret learning and mixed Nash equilibria: They do not
  mix},'' in \emph{Advances in Neural Information Processing Systems}, Virtual,
  2020, pp. 1380--1391.

\bibitem{daskalakis2018last}
C.~Daskalakis and I.~Panageas, ``{Last-iterate convergence: Zero-sum games and
  constrained min-max optimization},'' \emph{arXiv preprint arXiv:1807.04252},
  2018.

\bibitem{abernethy2019fictitious}
J.~Abernethy, K.~A. Lai, and A.~Wibisono, ``Last-iterate convergence rates for
  min-max optimization,'' \emph{arXiv preprint arXiv:1906.02027}, 2019.

\bibitem{golowich2020last}
N.~Golowich, S.~Pattathil, C.~Daskalakis, and A.~Ozdaglar, ``{Last iterate is
  slower than averaged iterate in smooth convex-concave saddle point
  problems},'' in \emph{Annual Conference on Learning Theory}, 2020, pp.
  1758--1784.

\bibitem{conitzer2011commitment}
V.~Conitzer and D.~Korzhyk, ``Commitment to correlated strategies,'' in
  \emph{AAAI Conference on Artificial Intelligence}, California, USA, 2011, pp.
  632--637.

\bibitem{bnnobrs1962partitioning}
J.~F. Benders, ``Partitioning procedures for solving mixed-variables
  programming problems,'' \emph{Numerische Mathematik}, vol.~4, no.~1, pp.
  238--252, 1962.

\bibitem{farkas1902theorie}
J.~Farkas, ``{Theorie der einfachen Ungleichungen},'' \emph{Journal f{\"u}r die
  reine und angewandte Mathematik}, vol. 1902, no. 124, pp. 1--27, 1902.

\bibitem{fischetti2009minimal}
M.~Fischetti, D.~Salvagnin, and A.~Zanette, ``{Minimal infeasible subsystems
  and Benders cuts}.''

\bibitem{gomory1958outline}
R.~E. Gomory, ``Outline of an algorithm for integer solutions to linear
  programs,'' \emph{Bulletin of the American Mathematical Society}, vol.~64,
  pp. 275--278, 1958.

\bibitem{land1960automatic}
A.~H. Land and A.~G. Doig, ``An automatic method of solving discrete
  programming problems,'' \emph{Econometrica}, vol.~28, no.~3, pp. 497--520,
  1960.

\bibitem{ruder2016overview}
S.~Ruder, ``An overview of gradient descent optimization algorithms,''
  \emph{arXiv preprint arXiv:1609.04747}, 2016.

\bibitem{gottipati2021genetic}
S.~Gottipati and P.~Paruchuri, ``{A genetic algorithm approach to compute mixed
  strategy solutions for general Stackelberg games},'' in \emph{IEEE Congress
  on Evolutionary Computation}, Krakow, Poland, 2021, pp. 1648--1655.

\bibitem{de2016machine}
G.~De~Nittis and F.~Trovo, ``{Machine learning techniques for Stackelberg
  security games: A survey},'' \emph{arXiv preprint arXiv:1609.09341}, 2016.

\bibitem{tran2021hamilton}
H.~V. Tran, \emph{Hamilton-Jacobi Equations: Theory and Applications}.\hskip
  1em plus 0.5em minus 0.4em\relax American Mathematical Soc., 2021, vol. 213.

\bibitem{li2020multiplayer}
M.~Li, J.~Qin, N.~M. Freris, and D.~W.~C. Ho, ``{Multi-player Stackelberg-Nash
  Game for nonlinear system via value iteration-based integral reinforcement
  learning},'' \emph{IEEE Transactions on Neural Networks and Learning
  Systems}, vol.~30, no.~4, pp. 1429--1440, 2022.

\bibitem{ontanon2013survey}
S.~Ontan{\'o}n, G.~Synnaeve, A.~Uriarte, F.~Richoux, D.~Churchill, and
  M.~Preuss, ``{A survey of real-time strategy game AI research and competition
  in StarCraft},'' \emph{IEEE Transactions on Computational Intelligence and AI
  in Games}, vol.~5, no.~4, pp. 293--311, 2013.

\bibitem{davidai2019politics}
S.~Davidai and M.~Ongis, ``{The politics of zero-sum thinking: The relationship
  between political ideology and the belief that life is a zero-sum game},''
  \emph{Science Advances}, vol.~5, no.~12, pp. 1--10, 2019.

\bibitem{von2020multiple}
A.~Von~Moll, E.~Garcia, D.~Casbeer, M.~Suresh, and S.~C. Swar,
  ``Multiple-pursuer, single-evader border defense differential game,''
  \emph{Journal of Aerospace Information Systems}, vol.~17, no.~8, pp.
  407--416, 2020.

\bibitem{gao2019communication}
X.~Gao, E.~Akyol, and T.~Basar, ``Communication scheduling and remote
  estimation with adversarial intervention,'' \emph{IEEE/CAA Journal of
  Automatica Sinica}, vol.~6, no.~1, pp. 32--44, 2019.

\bibitem{na2020theoretical}
X.~Na and D.~Cole, ``Theoretical and experimental investigation of driver
  noncooperative-game steering control behavior,'' \emph{IEEE/CAA Journal of
  Automatica Sinica}, vol.~8, no.~1, pp. 189--205, 2021.

\bibitem{albert2022homeland}
L.~A. Albert, A.~Nikolaev, and S.~H. Jacobson, ``Homeland security research
  opportunities,'' \emph{IISE Transactions}, pp. 1--23, 2022.

\bibitem{song2011mimo}
X.~Song, P.~Willett, S.~Zhou, and P.~B. Luh, ``{The MIMO radar and jammer
  games},'' \emph{IEEE Transactions on Signal Processing}, vol.~60, no.~2, pp.
  687--699, 2011.

\bibitem{li2022counterfactual}
H.~Li, Z.~Han, W.~Pu, L.~Liu, K.~Li, and B.~Jiu, ``Counterfactual regret
  minimization for anti-jamming game of frequency agile radar,'' \emph{arXiv
  preprint arXiv:2202.10049}, 2022.

\bibitem{bachmann2011game}
D.~J. Bachmann, R.~J. Evans, and B.~Moran, ``Game theoretic analysis of
  adaptive radar jamming,'' \emph{IEEE Transactions on Aerospace and Electronic
  Systems}, vol.~47, no.~2, pp. 1081--1100, 2011.

\bibitem{garcia2019strategies}
E.~Garcia, A.~Von~Moll, D.~W. Casbeer, and M.~Pachter, ``Strategies for
  defending a coastline against multiple attackers,'' in \emph{Proceedings of
  IEEE Conference on Decision and Control (CDC)}, Nice, France, 2019, pp.
  7319--7324.

\bibitem{lelis2021planning}
L.~H.~S. Lelis, ``{Planning algorithms for zero-sum games with exponential
  action spaces: A unifying perspective},'' in \emph{International Conference
  on International Joint Conferences on Artificial Intelligence}, 2021, pp.
  4892--4898.

\bibitem{liu2022learning}
Q.~Liu, Y.~Wang, and C.~Jin, ``{Learning Markov games with adversarial
  opponents: Efficient algorithms and fundamental limits},'' \emph{arXiv
  preprint arXiv:2203.06803}, 2022.

\bibitem{banik2021attack}
S.~Banik and S.~D. Bopardikar, ``Attack-resilient path planning using dynamic
  games with stopping states,'' \emph{IEEE Transactions on Robotics}, vol.~38,
  no.~1, pp. 25--41, 2021.

\bibitem{wang2017generative}
K.~Wang, C.~Gou, Y.~Duan, Y.~Lin, X.~Zheng, and F.-Y. Wang, ``{Generative
  adversarial networks: Introduction and outlook},'' \emph{IEEE/CAA Journal of
  Automatica Sinica}, vol.~4, no.~4, pp. 588--598, 2017.

\bibitem{lee2021last}
C.-W. Lee, C.~Kroer, and H.~Luo, ``Last-iterate convergence in extensive-form
  games,'' in \emph{Advances in Neural Information Processing Systems},
  vol.~34, 2021, pp. 1--13.

\bibitem{perolat2021poincare}
J.~Perolat, R.~Munos, J.-B. Lespiau, S.~Omidshafiei, M.~Rowland, P.~Ortega,
  N.~Burch, T.~Anthony, D.~Balduzzi, B.~De~Vylder \emph{et~al.}, ``{From
  Poincar{\'e} recurrence to convergence in imperfect information games:
  Finding equilibrium via regularization},'' in \emph{International Conference
  on Machine Learning}, 2021, pp. 8525--8535.

\bibitem{henderson2021cybered}
H.~Henderson, ``Cybered competition, cooperation, and conflict in a game of
  imperfect information,'' \emph{The Cyber Defense Review}, vol.~6, no.~3, pp.
  43--60, 2021.

\bibitem{costikyan2013uncertainty}
G.~Costikyan, \emph{Uncertainty in Games}.\hskip 1em plus 0.5em minus
  0.4em\relax MIT Press, 2013.

\bibitem{xu2021learning}
L.~Xu, ``Learning and planning under uncertainty for green security,'' in
  \emph{International Joint Conference on Artificial Intelligence}, 2021, pp.
  1--3.

\bibitem{kar2015game}
D.~Kar, F.~Fang, F.~Delle~Fave, N.~Sintov, and M.~Tambe, ``{A game of thrones:
  When human behavior models compete in repeated Stackelberg security games},''
  in \emph{Proceedings of the 2015 International Conference on Autonomous
  Agents and Multiagent Systems}, 2015, pp. 1381--1390.

\bibitem{caballero2021identifying}
W.~N. Caballero, B.~J. Lunday, and R.~P. Uber, ``Identifying behaviorally
  robust strategies for normal form games under varying forms of uncertainty,''
  \emph{European Journal of Operational Research}, vol. 288, no.~3, pp.
  971--982, 2021.

\bibitem{tsiotras2021bounded}
P.~Tsiotras, ``Bounded rationality in learning, perception, decision-making,
  and stochastic games,'' in \emph{Handbook of Reinforcement Learning and
  Control}, 2021, pp. 491--523.

\bibitem{platzer2015differential}
A.~Platzer, ``Differential game logic,'' \emph{ACM Transactions on
  Computational Logic}, vol.~17, no.~1, pp. 1--51, 2015.

\bibitem{iyermodeling}
M.~Iyer and B.~Gilby, ``Modeling an adversarial poacher-ranger hybrid game.''

\bibitem{brown2020combining}
N.~Brown, A.~Bakhtin, A.~Lerer, and Q.~Gong, ``Combining deep reinforcement
  learning and search for imperfect-information games,'' in \emph{Advances in
  Neural Information Processing Systems}, vol.~33, 2020, pp. 17\,057--17\,069.

\bibitem{li2020openholdem}
K.~Li, H.~Xu, M.~Zhang, E.~Zhao, Z.~Wu, J.~Xing, and K.~Huang, ``{OpenHoldem:
  An open toolkit for large-scale imperfect-information game research},''
  \emph{arXiv preprint arXiv:2012.06168}, 2020.

\bibitem{oh2021creating}
I.~Oh, S.~Rho, S.~Moon, S.~Son, H.~Lee, and J.~Chung, ``{Creating pro-level AI
  for a real-time fighting game using deep reinforcement learning},''
  \emph{IEEE Transactions on Games}, DOI: 10.1109/TG.2021.3049539, 2021.

\end{thebibliography}
\end{document}